\documentclass[12pt]{article}
\usepackage{amssymb,amsmath,epsfig}
\usepackage{graphicx}
\usepackage{subfig}
\usepackage{color}
\usepackage[utf8]{inputenc}
\usepackage{cite}
 \allowdisplaybreaks
\begin{document}
\title{\bf Energy Conditions in Extended $f(R,G,T)$ Gravity}

\author{M. Ilyas $^1$
\thanks{ilyas\_mia@yahoo.com}, Aftab Ahmad $^1$ \thanks{aftabahmad@gu.edu.pk},  Fawad Khan $^1$ \thanks{fawad.ccms@gmail.com} and
M. Wasif$^1$ \thanks{wasifkhawaja8@gmail.com}\\
$^{1}$Institute of Physics, Gomal University,\\
Dera Ismail Khan, 29220, KP, Pakistan\\}

\date{}

\maketitle
\begin{abstract}
In this paper, we {consider} the flat Friedmann–Lemaître–Robertson-Walker metric in the {presence} of perfect fluid models and extended $f(R,G,T)$ gravity {(where $R$ is the Ricci scalar, $G$ is the Gauss Bonnet invariant and $T$ stands for trace of energy momentum tensor)}. {In this context, we  assume some specific realistic $f(R,G,T)$ models configuration that could be used to explore the finite-time future singularities that arise in late-time cosmic accelerating phases}. {In this scenario, we choose} the most recent estimated values for the Hubble, deceleration, snap and jerk parameters to develop the viability and bounds on the models parameters induced by different energy conditions.
\end{abstract}

\section{Introduction}

Studies of Supernovae Type Ia, {cosmic microwave background (CMB)}, and other related phenomena have yielded a number of intriguing results \cite{ya1,ya1(b),ya1(c)}. In  domains of relativistic cosmology and astrophysics, {it provides a greater impact and open a new research platform}. These findings back up the theory that the universe's current expansion is speeding up. Observational data from satellites such as the Planck~\cite{ya2, Planck:2015xua, Ade:2015lrj}, the BICEP2 experiment~\cite{Ade:2014xna, Ade:2015tva, Array:2015xqh}, and the Wilkinson microwave anisotropy probe (WMAP)~\cite{Komatsu:2010fb, Hinshaw:2012aka}, {indicated that the universe contains: 68$\%$ dark energy (DE), 27$\%$ dark matter (DM) and 27$\%$ baryonic matter (BM)}. Many relativistic astrophysicists have proposed the idea of modified gravity theories (MGTs), which {were} derived via changing the $R$ in  conventional Einstein-Hilbert (EH) action. The formulations of this technique could be used as a starting point for investigating the causes of cosmic rapid expansion (see for example, ~\cite{R1,R2,R4,R5,R6,R7,R8,R9,R10,3,4,5,6,8}). Nojiri and Odintsov provided the first systematic results  of pulsating universe via $f(R)$ gravity \cite{n01}. {Some interesting results are obtained on the explanation of dark source terms on the dynamical evolution of stellar systems in Einstein-cosmological constant (-$\Lambda$) \cite{zs1}, the modified theories of gravity {MGTs} i.e.,  the $f(R)$ \cite{z1fr(a)}, the $f(R,T)$ \cite{z2frt} and $f(T,R,R_{\mu\nu}T^{\mu\nu})$ gravity} \cite{z3frtrmn}.  The Gauss Bonnet (GB) gravity is one of the sophisticated MGTs, which has received novel attraction \cite{9,10,11,12,13,14,15,16,17,18,19} and is named as $f(G)$ gravity, where $G = R^2 - 4{R_{\mu \nu }}{R^{\mu \nu }} + {R_{\mu \nu\alpha \beta }}{R^{\mu \nu \alpha \beta }}$ is a topological Lorentz invariant. { For many different scenarios, this MGT could be beneficial for researching the inflationary era, transitions between acceleration and deceleration zones, passing solar system trials, and traversing phantom separation walls models with the $f(G)$ \cite{11,12}}. The GB gravity is likewise shown to be less limited than the $f(R)$ gravity~\cite{13}. As an alternative to dark energy, $f(G)$ gravity provides a systematic method for studying many cosmic issues~\cite{14}. It has the potential to be extremely useful inside the study of restricted period future discontinuities and even the pace of cosmos over long periods of time~\cite{15,16}. Similarly, multiple existing theories in $f(G)$ gravity are helpful to tell the cosmic acceleration followed by the matter era \cite{13,14}. Various $f(G)$ modifications were devised to overcome unique solar framework {constraints}~\cite{13,14}, that are explained in \cite{17}. The consideration of energy circumstances may also yield further bounds on $f(G)$ models (ECs)~\cite{18,19,20}. Nojiri \emph{et al.} \cite{bin4} studied key cosmic problems like inflation, late-time acceleration, and {bouncing cosmology}, and came to the conclusion that certain customized theories of gravity could be considered a reasonable numerical framework for understanding our current universe's.\\

{The energy conditions are the basic ingredients to understand  the black-hole thermodynamics theorem, serve as a backbone for deep learning of the singularity theorem}. For weak energy (WE) and strong energy (SE) situations, the Hawking-Penrose singularity theorem is beneficial, whereas the black hole second law of thermodynamics is useful for null energy (NE). To discuss the consistency of different types of ECs, Raychaudhuri equation might used \cite{21,22,23,23a,23b,23c,23d}. The energy condition was described in the existing literature employing classical ECs of general relativity (GR) such as the phantom fields potential  \cite{24}, the background of extended world\cite{25,26,27,28,29,30} and the deceleration parameters' pattern moving \cite{31,32}. The various expressions for ECs are generated using these tools and $f(R)$ gravity~\cite{33}. {Most of the relevant experts in the field have raised some (cosmological) concerns about $f(R)$ gravity \cite{34,35,36}}. {Garc\'{i}a \emph{et al.}  developed the generic ontology of ECs in $f(G)$ gravity\cite{37}, also in Ref.~\cite{gar1}, they have presented few $f(G)$ models and explored ECs to assess their viability epochs}. By researching the dynamical behaviour of Weak energy conditions (WECs), Nojiri \emph{et al.} \cite{15} {established some} plausible $f(G)$ models. Sadeghi \emph{et al.} \cite{sad1} looked at certain $f(G)$ gravity models that obey WECs and {strong energy conditions} (SECs) in a time when the late-time de-Sitter solution was stable. Banijamali
\emph{et al.}~\cite{bani1} studied the WECs distribution for a class of consistent $f(G)$ models and {predicted that} a power law model of kind $f(G)=\epsilon G^n$ would satisfy  the WECs when $\epsilon<0$ was set. The various aspects of charged compact stars is analyzed in Ref.~\cite{chargemia}. \\

{The MTGs, which represents gravitational interactions differently than the more well-known theory of general relativity, has garnered a lot of interest.}\\

We took some of the estimated values of the Hubble, deceleration, jerk, and snap model parameters in this paper. The bounds on the model parameters of $f(R,G,T)$ gravity are derived from ECs, as indicated in the Ref.~\cite{13}. {With the help of using several plots, we demonstrated the considered models in extended $f(R,G,T)$ gravity may satisfy the different $ECs$ in a particular region, which is required for investigating the stability of late time de-Sitter solutions.} The following section of this research work is organized as: In section \textbf{2}, { We present a quick overview of $f(R,G,T)$ field equations as well as a effective version of $ECs$}. {In  section \textbf{3}, we  analyze the viability epochs of ECs, we look at some viable models in extended $f(R,G,T)$ gravity}. {The final section~\textbf{4}, we summarizes the outcome of our work and  draw conclusions.}
\section{Extended $f(R,G,T)$ Gravity}
The Hilbert-Einstein action for $f(G,R,T)$ is written as \cite{ilyas4}
\begin{equation}\label{1}
S_{f(T,R,G)} = \frac{1}{{2{\kappa ^2}}}\int {\sqrt { - g} \left[ {f(R,G,T) + {\Psi}} \right]} {d^4}x,
\end{equation}
where $f(R,G,T)$ denotes the arbitrary function of $R$, $G$ and $T$ while $\Psi$ is matter-Lagrangian, $g=|g_{\mu\nu}|$ and $\kappa^{2}=8\pi G$ with $c=1$. The matter's stress-energy tensor is defined as
\begin{equation}\label{2}
{T_{\alpha \beta }} =  - \frac{2}{{\sqrt { - g} }}\frac{{\delta \left[ {\sqrt { - g} {\Psi}} \right]}}{{\delta {g^{\alpha \beta }}}}.
\end{equation}
The trace of the stress-energy tensor is $T=g_{\mu\nu}T^{\mu\nu}$. We suppose that the matter Lagrangian $\Psi$  is simply dependent on the metric tensor $g_{\mu\nu}$ in this case, therefore we get
\begin{equation}\label{3}
T_{\mu\nu}=g_{\mu\nu}\Psi-2~\frac{\partial \Psi}{\partial
g^{\mu\nu}}.
\end{equation}
The field equations of gravity $f(R,G,T)$ are obtained as follows
\begin{align}\label{13}\nonumber
&\left[ {{R_{\alpha \beta }} + {g_{\alpha \beta }}{\Box} - {\nabla _\alpha }{\nabla _\beta }} \right]{f_R} - \frac{1}{2}f{g_{\alpha \beta }} + (2R{R_{\alpha \beta }} - 4R_\alpha ^\xi {R_{\xi \beta }} - 4{R_{\alpha \xi \beta \eta }}{R^{\xi \eta }}\\\nonumber
& + 2R_\alpha ^{\xi \eta \lambda }{R_{\beta \xi \eta \lambda }}){f_G} + (2R{g_{\alpha \beta }}{\Box} - 2R{\nabla _\alpha }{\nabla _\beta } - 4{g_{\alpha \beta }}{R^{\xi \eta }}{\nabla _\xi }{\nabla _\eta } - 4{R_{\alpha \beta }}{\Box}\\
& + 4R_\alpha ^\xi {\nabla _\beta }{\nabla _\xi } + 4R_\beta ^\xi {\nabla _\alpha }{\nabla _\xi } + 4{R_{\alpha \xi \beta \eta }}{\nabla ^\xi }{\nabla ^\eta }){f_G} = {\kappa ^2}{T_{\alpha \beta }} - ({\Theta _{\alpha \beta }} + {T_{\alpha \beta }}){f_{\cal T}},
\end{align}
where the d'Alembert operator  is $\Box=\nabla_{\mu}\nabla^{\mu}$. We may get the trace of the above field equation  (\ref{13}) by multiplying both sides by $g^{\mu\nu}$.
\begin{equation}
\left[ {3{\Box} + R} \right]{f_R} + (2R{\Box} - 4{R^{\alpha \beta }}{\nabla _\alpha }{\nabla _\beta }- 2G){f_G} - 2f = {\kappa ^2}T - (T+ \Theta ){f_{\cal T}},
\end{equation}
where $\Theta=\Theta_{\mu\nu}g^{\mu\nu}$. Taking covariant divergence of equation (\ref{13}), we get
\begin{equation}\label{15}
{\nabla ^\alpha }{T_{\alpha \beta }} = \frac{{{f_T}}}{{{\kappa ^2} - {f_T}}}\left[ {{\nabla ^\alpha }{\Theta _{\alpha \beta }} + \left\{ {{\Theta _{\alpha \beta }} + {T_{\alpha \beta }}} \right\}{\nabla ^\alpha }\ln {f_T} - \frac{1}{2}{g_{\alpha \beta }}{\nabla ^\alpha }T} \right].
\end{equation}
We can see that the above equation is not affected by $f_{R}$ and $f_{G}$.  We could also get
To obtain a useful expression for $\Theta_{\alpha\beta}$, we
differentiate Eq.(\ref{3}) with respect to metric tensor
\begin{equation}\label{6a}
\frac{\delta T_{\alpha\beta}}{\delta g^{\xi\eta}}=\frac{\delta
g_{\alpha\beta}}{\delta g^{\xi\eta}}\Psi+g_{\alpha\beta}
\frac{\partial\Psi}{\partial
g^{\xi\eta}}-2\frac{\partial^2\Psi}{\partial
g^{\xi\eta}\partial g^{\alpha\beta}}.
\end{equation}
Using the relations
\begin{equation}\nonumber
\frac{\delta g_{\alpha\beta}}{\delta
g^{\xi\eta}}=-g_{\alpha\mu}g_{\beta\nu}\delta_{\xi\eta}^{\mu\nu},\quad
\delta_{\xi\eta}^{\mu\nu}=\frac{\delta g^{\mu\nu}}{\delta
g^{\xi\eta}},
\end{equation}
where $\delta_{\xi\eta}^{\mu\nu}$ is the generalized Kronecker
symbol and Eq.(\ref{6a}) in following equation,
$$\quad\Theta_{\alpha\beta}=g^{\xi\eta}\frac{\delta
T_{\xi\eta}}{\delta g_{\alpha\beta}}.$$
we obtain
\begin{equation}\label{16}
{\Theta _{\alpha \beta }} = {g_{\alpha \beta }}{\Psi} - 2{T_{\alpha \beta }} - 2{g^{\eta \lambda }}\frac{{{\partial ^2}{\Psi}}}{{\partial {g^{\alpha \beta }}\partial {g^{\eta \lambda }}}}
\end{equation}
If $\Psi$ is known then  $\Theta_{\mu\nu}$ can be found. Energy momentum tensor for ideal fluid is
\begin{equation}
{T_{\eta \lambda }} = \left[ {\rho  + P} \right]{u_\eta }{u_\lambda } + P{g_{\eta \lambda }}
\end{equation}
where  $P$ and $\rho$ are respectively  pressure and  energy density of perfect fluid. The four velocity $u_{\mu}$ satisfies $u_{\mu}u^{\mu}=-1$ and $u^{\mu}\nabla _{\nu}u_{\mu}=0$. Currently we assume that  Lagrangian for matter is $\Psi=p$. As a result, the equation (\ref{16}) reduced as
\begin{equation}
{\Theta _{\eta \lambda }} =  - 2{T_{\eta \lambda }} + P{g_{\eta \lambda }}
\end{equation}
The line element is used to expect a flat FLRW version of the universe,
\begin{equation}
d{s^2} = {a^2}(t)\left( {d{r^2} + {r^2}d{\theta ^2} + {r^2}si{n^2}\theta d{\phi ^2}} \right) - d{t^2},
\end{equation}
where $a(t)$ is the scale factor, it may derive as
\begin{equation}
R = 6\dot H + 12{H^2},~~G = 24{H^2}\dot H + 24{H^4},
\end{equation}
where $H=\dot{a}/a$  {represents} the Hubble parameter and $dot$ denotes the derivative with respect to cosmic time $t$. We get the trace of $T_{\mu \nu}$ is $T=[3P-\rho]$ and $\Theta=2[\rho-P]$ using the above metric. So $T+\Theta=2(\rho+P)$. We derive the equation of non-conservation by(\ref{15}).
\begin{equation}\label{21}
\dot \rho  + 3\left( {\rho  + P} \right)H = \left( {\frac{1}{2}\dot T - \dot P} \right){f_T} - \left( {\rho  + P} \right){\dot f_T}.
\end{equation}
The basic conservation equation $\nabla^{\alpha}T_{\alpha\beta}=0$ for a perfect fluid.
\begin{equation}\label{22}
\dot{\rho}+3(\rho+P)H=0
\end{equation}
As a result of equation (\ref{21}), we get
\begin{equation}
\left[ {\dot \rho  - \dot P} \right]{f_T} =- 2\left( {\rho  + P} \right){\dot f_T},
\end{equation}
The field equations for $f(R,G,T)$ gravity are obtained with equation (\ref{13}).
\begin{align}\label{24}\nonumber
3{H^2} = &{f_R}^{ - 1}\left[ {{\kappa ^2}\rho } \right. + \frac{1}{2}(R{f_R} - f) - 3H{{\dot f}_R} + 12{H^2}\dot H{f_G}\\
& + 12{H^4}{f_G} - 12{H^3}{{\dot f}_G} + (\rho  + \left. {P){f_T}} \right],
\end{align}
\begin{align}\label{25}\nonumber
2\dot H + 3{H^2} = & - {f_R}^{ - 1}\left[ {{\kappa ^2}\rho  + } \right.\frac{1}{2}(R{f_R} - f) - 3H{{\dot f}_R} + 12{H^2}\dot H{f_G}\\
& + 12{H^4}{f_G} - 12{H^3}{{\dot f}_G} + (\rho \left. { + P){f_T}} \right].
\end{align}
In Einstein's standard field equations, the two upper field equations can be expressed as
\begin{equation}
\kappa^{2}\rho_{eff}=3H^{2}~~and~~
-\kappa^{2}P_{eff}=2\dot{H}+3H^{2}.
\end{equation}
where
\begin{align}\label{f1}\nonumber
{\rho _{eff}} = &{\kappa ^{ - 2}}{f_R}^{ - 1}\left[ {{\kappa ^2}\rho } \right. - \frac{1}{2}(f - R{f_R}) - 3H{{\dot f}_R} + 12{H^2}\dot H{f_G}\\
& + 12{H^4}{f_G} - \left. {12{H^3}{{\dot f}_G} + (\rho  + P){f_{\cal T}}} \right],
\end{align}
\begin{align}\label{f2}\nonumber
{P_{eff}} = &{\kappa ^{ - 2}}{f_R}^{ - 1}\left[ {{\kappa ^2}P} \right. + \frac{1}{2}(f - R{f_R}) + 2H{{\dot f}_R} + {{\ddot f}_R} - (12{H^2}\dot H\\
& + 12{H^4}){f_G} + 8H(\dot H + {H^2}){{\dot f}_G} + 4\left. {{H^2}{{\ddot f}_G}} \right].
\end{align}
Further, the first and second derivatives of  $f_{R}$ and $f_{G}$  with respect to $t$ are now shown as
\begin{align}\label{31}\nonumber
{{\dot f}_R}& = \dot R{f_{RR}} + \dot G{f_{RG}} + \mathop T\limits^. {f_{RT}},{{\dot f}_G} = \dot R{f_{RG}} + \dot G{f_{GG}} + \dot T{f_{GT}},\\\nonumber
{{\ddot f}_R} =& \ddot R{f_{RR}} + \ddot G{f_{RG}} + \ddot T{f_{RT}} + {{\dot R}^2}{f_{RRR}} + 2\dot R\dot G{f_{RRG}}\\\nonumber
& + 2\dot R\dot T{f_{RRT}} + {{\dot G}^2}{f_{RGG}} + 2\dot G\dot T{f_{RGT}} + {{\dot T}^2}{f_{RTT}},\\\nonumber
{{\ddot f}_G} =& \ddot R{f_{RG}} + \ddot G{f_{GG}} + \ddot T{f_{GT}} + {{\dot R}^2}{f_{RRG}} + 2\dot R\dot G{f_{RGG}}\\
& + 2\dot R\dot T{f_{RGT}} + {{\dot G}^2}{f_{GGG}} + 2\dot G\mathop T\limits^. {f_{GGT}} + {{\dot T}^2}{f_{GTT}}.
\end{align}
The field equations (\ref{24}) and (\ref{25}) are reduced to the following equation using (\ref{31}),
\begin{align}\label{32}\nonumber
{\kappa ^2}\rho  &- \frac{1}{2}f + \left[ {\frac{R}{2} - 3{H^2}} \right]{f_R} + \left[ {12{H^2}\dot H + 12{H^4}} \right]{f_G} - 3H\dot R{f_{RR}} + (\rho  + P){f_T}\\
& - \left[ {12{H^3}\dot R + 3H\dot G} \right]{f_{RG}} - 12{H^3}\dot G{f_{GG}} - 3H\dot T{f_{RT}} - 12{H^3}\dot T{f_{GT}} = 0,
\end{align}
and
\begin{align}\label{33}\nonumber
&{\kappa ^2}P + \frac{1}{2}f + \frac{1}{2}\left[ {4\dot H + 6{H^2} - R} \right]{f_R} - \left\{ {12{H^4} + 12{H^2}\dot H} \right\}{f_G} + \left\{ {\ddot R + 2H\dot R} \right\}{f_{RR}}\\\nonumber
& + \left\{ {\ddot T + 2H\dot T} \right\}{f_{RT}} + \left[ {\left\{ {8{H^3} + 8H\dot H} \right\}\dot R + 4{H^2}\ddot R + 2H\dot G + \ddot G} \right]{f_{RG}} + 2\dot R\dot T{f_{RRT}}\\\nonumber
& + \left[ {\left\{ {8{H^3} + 8H\dot H} \right\}\dot G + 4{H^2}\ddot G} \right]{f_{GG}}\left[ {\left\{ {8{H^3} + 8H\dot H} \right\}\dot T + 4{H^2}\ddot T} \right]{f_{GT}} + {{\dot R}^2}{f_{RRR}}\\\nonumber
& + \left\{ {4{H^2}{{\dot R}^2} + 2\dot R\dot G} \right\}{f_{RRG}} + \left\{ {8{H^2}\dot R\dot G + {{\dot G}^2}} \right\}{f_{RGG}} + \left\{ {8{H^2}\dot R\dot T + 2\dot G\dot T} \right\}{f_{RGT}}\\
& + {{\dot T}^2}{f_{RTT}} + 4{H^2}{{\dot G}^2}{f_{GGG}} + 8{H^2}\dot G\dot T{f_{GGT}} + 4{H^2}{{\dot T}^2}{f_{GTT}} = 0.
\end{align}

{In the upcoming sections}, we build the $f(R,G,\mathcal{T})$ for de-Sitter, power law, and future singularity expansion models in terms of $R,~G$ and $\mathcal{T}$.

\section{Energy Conditions}

The ECs are the fundamental and essential equipment for black holes, wormholes (WHs) and other physical scenarios. The breaking of these limitations may be useful in determining the stability of WHs. Because the sphere equations fluctuates from the Einstein equations {in the situation when researching ECs in MGTs is significantly different}. The ECs in GR are derived {by} relating $R_{\mu\nu}$ with regular energy momentum tensor.  It is necessary to understand how to tie $R_{\mu\nu}$ to the powerful styles of the energy momentum tensor, with a view to result in the related ECs. Raychaudhuri's equation for the growth nature produces those ECs. The NEC and WEC in MGTs (with powerful energy density and pressure) are defined as follows:
\begin{align}
&\textrm{NEC}: {\rho^{eff}} + {P^{eff}} \ge 0,\\\nonumber
&\textrm{WEC}:
{\rho^{eff}} \ge 0  \textrm{ and } {\rho^{eff}} + {P^{eff}} \ge 0,
\end{align}
while the DEC  and the SEC provide
\begin{align}
&\textrm{SEC}: {\rho^{eff}}+ 3{P^{eff}} \ge 0 \textrm{ and }  {\rho^{eff}} +
{P^{eff}} \ge 0,\\\nonumber
&\textrm{DEC}: {\rho^{eff}} \ge 0  \textrm{ and }  {\rho^{eff}} \pm
{P^{eff}} \ge 0.
\end{align}
We can see that ECs would place some restrictions on the parameters used in the construction of $f(G)$ models \cite{37}. The general expressions of the NEC, WEC, SEC, and DEC for extended $f(R,G,T)$ gravity theory are:\\

$\bullet$ \textbf{NEC:}
\begin{align}\nonumber
{\rho _{eff}} + {P_{eff}} =& {\kappa ^{ - 2}}{f_R}^{ - 1}\left[ {{\kappa ^2}\left\{ {\rho  + P} \right\}} \right. - H{{\dot f}_R} + {{\ddot f}_R} + \left\{ {8H\dot H - 4{H^3}} \right\}{{\dot f}_G}\\
& + 4{H^2}{{\ddot f}_G} + \left. {\left\{ {\rho  + P} \right\}{f_T}} \right] \ge 0,
\end{align}

$\bullet$ \textbf{WEC:}
\begin{align}\nonumber
{\rho _{eff}} &= {\kappa ^{ - 2}}{f_R}^{ - 1}\left[ {{\kappa ^2}\rho } \right. + \frac{1}{2}\left\{ {12{H^2}\dot H + 12{H^4}} \right\} + \left\{ {12{H^2}\dot H + 12{H^4}} \right\}{f_G}\\
& - 3H{{\dot f}_R} - 12{H^3}\left. {{{\dot f}_G} + \left\{ {\rho  + P} \right\}{f_T}} \right] \ge 0.
\end{align}
\begin{align}\nonumber
{\rho _{eff}} + {P_{eff}} =& {\kappa ^{ - 2}}{f_R}^{ - 1}\left[ {{\kappa ^2}\left\{ {\rho  + P} \right\}} \right. - H{{\dot f}_R} + {{\ddot f}_R} + \left\{ {8H\dot H - 4{H^3}} \right\}{{\dot f}_G}\\
& + 4{H^2}{{\ddot f}_G} + \left. {\left\{ {\rho  + P} \right\}{f_T}} \right] \ge 0.
\end{align}

$\bullet$ \textbf{SEC:}
\begin{align}\nonumber
{\rho _{eff}} + 3{P_{eff}} &= {\kappa ^{ - 2}}{f_R}^{ - 1}\left[ {{\kappa ^2}\left\{ {\rho  + 3P} \right\} + 3H{{\dot f}_R} + 3{{\ddot f}_R}} \right. - \left\{ {R{f_R} - f} \right\} + 12{H^2}{{\dot f}_G}\\
&\left. { - \left\{ {24{H^2}\dot H + 24{H^4}} \right\}{f_G} + \left\{ {24H\dot H + 12{H^3}} \right\}{{\dot f}_G} + \left\{ {\rho  + P} \right\}{f_T}}\right]\ge0.
\end{align}

\begin{align}\nonumber
{\rho _{eff}} + {P_{eff}} =& {\kappa ^{ - 2}}{f_R}^{ - 1}\left[ {{\kappa ^2}\left\{ {\rho  + P} \right\}} \right. - H{{\dot f}_R} + {{\ddot f}_R} + \left\{ {8H\dot H - 4{H^3}} \right\}{{\dot f}_G}\\
& + 4{H^2}{{\ddot f}_G} + \left. {\left\{ {\rho  + P} \right\}{f_T}} \right] \ge 0.
\end{align}

$\bullet$ \textbf{DEC:}
\begin{align}\nonumber
{\rho _{eff}} &= {\kappa ^{ - 2}}{f_R}^{ - 1}\left[ {{\kappa ^2}\rho } \right. + \frac{1}{2}\left\{ {12{H^2}\dot H + 12{H^4}} \right\} + \left\{ {12{H^2}\dot H + 12{H^4}} \right\}{f_G}\\
& - 3H{{\dot f}_R} - 12{H^3}\left. {{{\dot f}_G} + \left\{ {\rho  + P} \right\}{f_T}} \right] \ge 0,
\end{align}
\begin{align}\nonumber
{\rho _{eff}} + {P_{eff}} =& {\kappa ^{ - 2}}{f_R}^{ - 1}\left[ {{\kappa ^2}\left\{ {\rho  + P} \right\}} \right. - H{{\dot f}_R} + {{\ddot f}_R} + \left\{ {8H\dot H - 4{H^3}} \right\}{{\dot f}_G}\\
& + 4{H^2}{{\ddot f}_G} + \left. {\left\{ {\rho  + P} \right\}{f_T}} \right] \ge 0,
\end{align}
and
\begin{align}\nonumber
{\rho _{eff}} - {P_{eff}} &= \frac{1}{{{\kappa ^2}{f_R}}}\left[ {{\kappa ^2}\left\{ {\rho  - P} \right\} + \left\{ {R{f_R} - f} \right\} - 5H{{\dot f}_R} - {{\ddot f}_R}} \right. - 4{H^2}{{\ddot f}_G}\\
&\left. { + \left\{ {24{H^2}\dot H + 24{H^4}} \right\}{f_G} - \left\{ {8H\dot H + 20{H^3}} \right\}{{\dot f}_G} + \left\{ {\rho  + P} \right\}{f_T}} \right] \ge 0
\end{align}
{It has been made apparent that the derivative of position four vector is called four- velocity, and its double derivative is called four-acceleration. Additionally, its third and fourth derivatives yield the jerk and snap parameters, respectively. As shown below, the Hubble parameter for a FLRW metric including perfect matter.}

\begin{align}\label{z3}
H=\frac{{a'}}{a},
\end{align}
whereas the jerk $j$, snap $s$, and deceleration $q$ factors are
\begin{align}\label{z4}
q =-\frac{1}{{{H^2}}}\frac{{a''}}{a},\quad j =
\frac{1}{{{H^3}}}\frac{{a'''}}{a}, \quad s =
\frac{1}{{{H^4}}}\frac{{{a''''}}}{a}.
\end{align}
The derivatives of Hubble parameters can be calculated
using these parameters.
\begin{align}\label{z5}
H'=-[q+1]H{^2},~~{H'''} = [- 2j- 5q+s-3]H{^4},~~ H''=[j + 3q+ 2]H{^3}.
\end{align}
Here, prime denotes the derivative with respect to $t$. From Eqs.~(\ref{f1})-(\ref{f2}), which can be rewritten as

\begin{align}\label{effrolast}\nonumber
\rho eff = &\frac{1}{{{\kappa ^2}}}{f_R}( - 12{H^3}(G'{f_{GG}} + R'{f_{RG}} + T'{f_{GT}}) - 3H(G'{f_{RG}}\\\nonumber
 +& R'{f_{RR}} + T'{f_{RT}}) + 12{H^2}\left( {{H^2} - {H^2}\left( {q + 1} \right)} \right){f_G} + \left( {p + \rho } \right){f_T}\\
 +& \frac{1}{2}(R{f_R} - f) + {\kappa ^2}\rho ),
\end{align}

\begin{align}\label{effplast}\nonumber
Peff = &\frac{1}{{{\kappa ^2}}}{f_R}({G^{\prime \prime }}{f_{RG}} + 8H({H^2} - {H^2}\left( {q + 1} \right))(G'{f_{GG}} + R'{f_{RG}}\\\nonumber
 +& T'{f_{GT}}) + 2H(G'{f_{RG}} + R'{f_{RR}} + T'{f_{RT}}) + T'(G'{f_{RGT}} + R'{f_{RRT}}\\\nonumber
 +& T'{f_{RTT}}) + G'(G'{f_{RGG}} + R'{f_{RRG}} + T'{f_{RGT}}) + R'(G'{f_{RRG}} + R'{f_{RRR}}\\\nonumber
 +& T'{f_{RRT}}) + 4{H^2}({G^{\prime \prime }}{f_{GG}} + T'(G'{f_{GGT}} + R'{f_{RGT}} + T'{f_{GTT}}) + G'(G'{f_{GGG}}\\\nonumber
 +& R'{f_{RGG}} + T'{f_{GGT}}) + R'(G'{f_{RGG}} + R'{f_{RRG}} + T'{f_{RGT}}) + {R^{\prime \prime }}{f_{RG}} + {T^{\prime \prime }}{f_{GT}})\\
 -& 12{H^2}({H^2} - {H^2}(q + 1)){f_G} + {R^{\prime \prime }}{f_{RR}} + {T^{\prime \prime }}{f_{RT}} + \frac{1}{2}\left( {f - R{f_R}} \right) + {\kappa ^2}p),
\end{align}
and
\begin{align}\label{effpplast}\nonumber
\rho eff + Peff =& \frac{1}{{{\kappa ^2}}}{f_R}({G^{\prime \prime }}{f_{RG}} - 4{H^3}\left( {2q + 3} \right)(G'{f_{GG}} + R'{f_{RG}} + T'{f_{GT}})\\\nonumber
 -& H(G'{f_{RG}} + R'{f_{RR}} + T'{f_{RT}}) + 2T'(G'{f_{RGT}} + R'{f_{RRT}})\\\nonumber
 +& 2G'R'{f_{RRG}} + {{G'}^2}{f_{RGG}} + 4{H^2}({G^{\prime \prime }}{f_{GG}} + 2T'(G'{f_{GGT}}\\\nonumber
 +& R'{f_{RGT}}) + 2G'R'{f_{RGG}} + {{G'}^2}{f_{GGG}} + {R^{\prime \prime }}{f_{RG}}\\\nonumber
 +& {{R'}^2}{f_{RRG}} + {T^{\prime \prime }}{f_{GT}} + {{T'}^2}{f_{GTT}}) + (p + \rho )\\
&({f_T} + {\kappa ^2}) + {R^{\prime \prime }}{f_{RR}} + {{R'}^2}{f_{RRR}} + {T^{\prime \prime }}{f_{RT}} + {{T'}^2}{f_{RTT}}).
\end{align}

\section{Some Viable Models}

In this part, we'll look at the impacts of several $f(R,G,T)$ models on the formulations and behavior of ECs with $\rho = a_0 e^{(-3t H )}$ in background.
For Hubble, decelerate, snap, and jerk factors, we  can use corresponding quantitative values.
in the following calculation~\cite{38}
$$H=0.718,~~q=-0.64,~~j=1.02,~~s=-0.39.$$
Those subcategories that follow will show us how to set up various FLRW model configurations managed by some specific $f(R,G,T)$ frameworks.

\subsection{Model 1}

\begin{figure} \centering
\epsfig{file=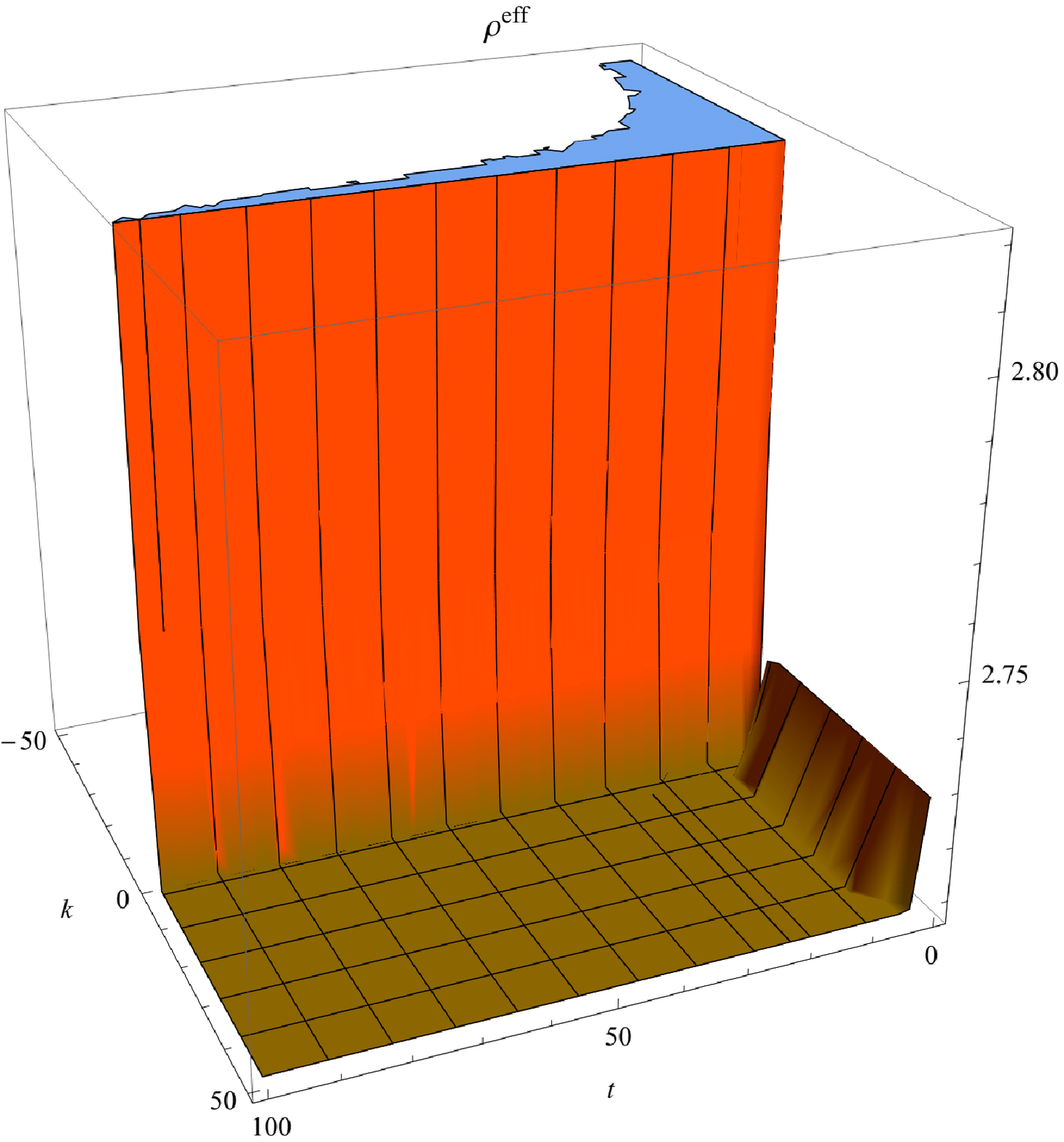,width=.48\linewidth}
\epsfig{file=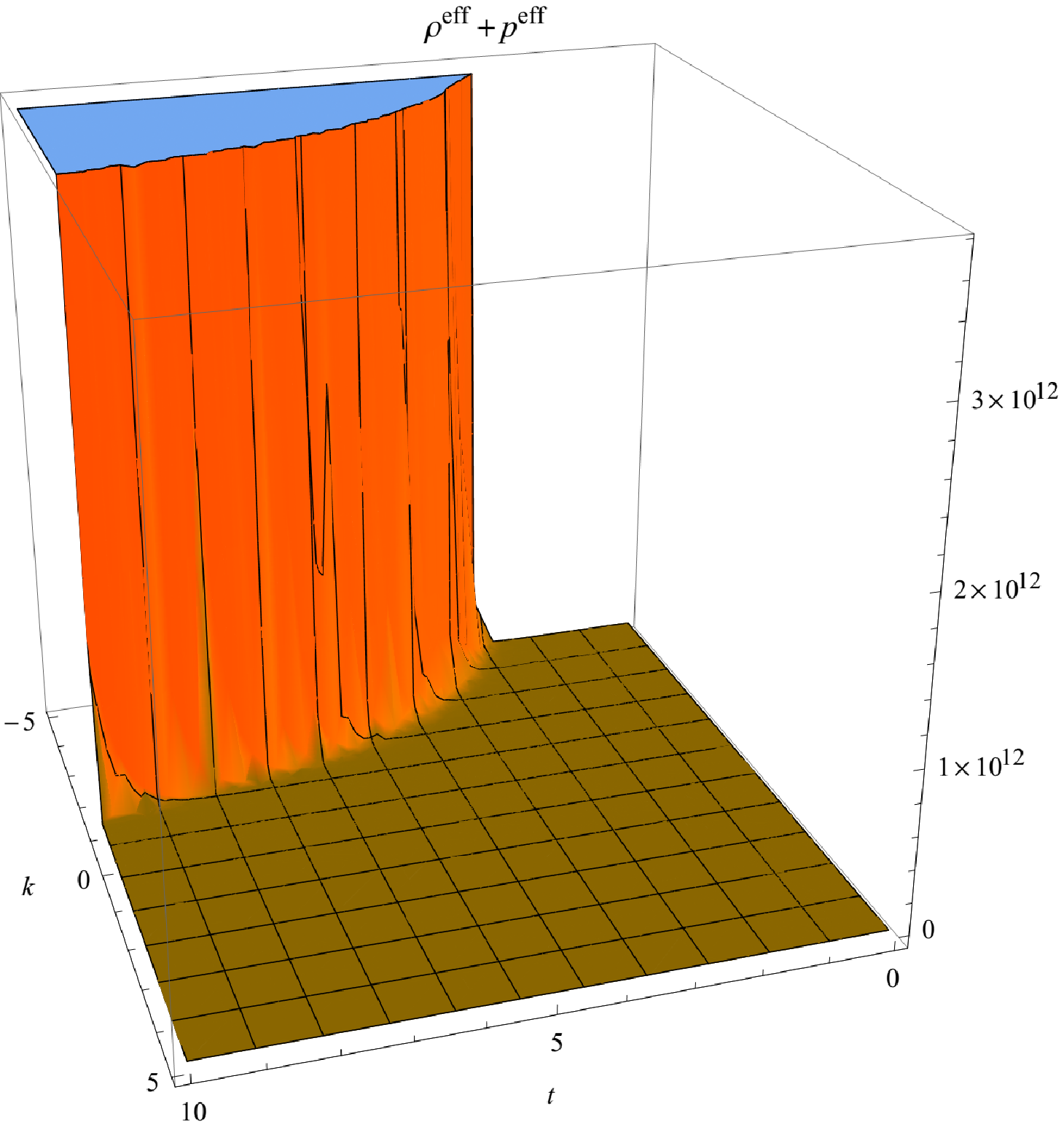,width=.48\linewidth} \caption{WEC plots for model 1 mentioned in Eq.(\ref{model1}) in extended $f(R,G,T)$ gravity. Left plot shows the behavior of $\rho^{eff}$ and right plot show the behavior of $\rho^{eff}+ P^{eff}$ with respect to $t$ and $k$ for $\alpha=\beta=1$, respectively.} \label{1f1}
\end{figure}

To begin, we'll use a framework that includes both power law and logarithm functions corrections to $f(G)$ \cite{39}, our inspiration {based on} the following model as
\begin{equation}\label{model1}
f(R,G,T) =R+R^2 +\alpha {G^n} + \beta G\log[G]+\gamma T^k,
\end{equation}
where $\alpha,~n$, $\beta$ and $k$ are parameters. The data supplied through the same cosmographic parameters \cite{40} is found to be in
agreement with the dynamics presented by this model. Substituting this into Eq.~(\ref{effrolast}), we get
the effective energy density as
\begin{align}\label{rom1}\nonumber
{\rho ^{eff}} =& \frac{1}{{{\kappa ^2}\left( {1 + 2R} \right)}}[12{H^2}(\beta  + n\alpha {G^{ - 1 + n}} + \beta \log \left[ G \right])\\\nonumber
&({H^2} - {H^2}(1 + q)) + \frac{1}{2}( - \alpha {G^n} - \beta G\log \left[ G \right] - R - {R^2}\\\nonumber
& + R(1 + 2R) - \gamma {T^k}) + {\kappa ^2}\rho  + k\gamma {T^{ - 1 + k}}\left( {p + \rho } \right) - 12\\
&\left\{ {\frac{\beta }{G} + ( - 1 + n)n\alpha {G^{ - 2 + n}}} \right\}{H^3}G' - 6HR']\geq 0.
\end{align}
When applying Eqs. (\ref{effpplast}), the total of effective pressure energy density can be deduced in the following way:
\begin{align}\label{roppm1}\nonumber
{\rho ^{eff}} + {P^{eff}} &= \frac{1}{{\left( {{\kappa ^2}{G^3}\left( {1 + 2R} \right)T} \right)}}[ - 4\beta G{H^2}T{{G'}^2} + 4n\\\nonumber
&\left( {{n^2} - 3n + 2} \right)\alpha {G^n}{H^2}T{{G'}^2} - 4\beta {G^2}{H^2}T(H(3 + 2q)\\\nonumber
&G' - {G^{\prime \prime }}) - 4\left( {n - 1} \right)n\alpha {G^{1 + n}}{H^2}T(H(3 + 2q)G' - {G^{\prime \prime }})\\
& + {G^3}(p({\kappa ^2}T + k\gamma {T^k}) + k\gamma {T^k}\rho  + T({\kappa ^2}\rho  - 2HR' + 2{R^{\prime \prime }}))]\geq 0.
\end{align}
It is a difficult to obtain an exact solution for the parameters  $\alpha$, $\beta$ $n$ and $k$ using the aforementioned two inequalities (\ref{rom1})
and (\ref{roppm1}). To accomplish this, we'll use a particular amount of $\alpha=\beta=1$ and plot $\rho^{\textrm{eff}}$ and $\rho^{\textrm{eff}}+ P^{\textrm{eff}}$ as  functions of $k$ and $n$ as  illustrated in fig.\ref{1f1}. Here we consider $0<t<10$, $-5<k<5$ and $-5<\alpha<5$. The validity of WECs is demonstrated in Fig. \ref{1f1}.\\

\begin{figure} \centering
\epsfig{file=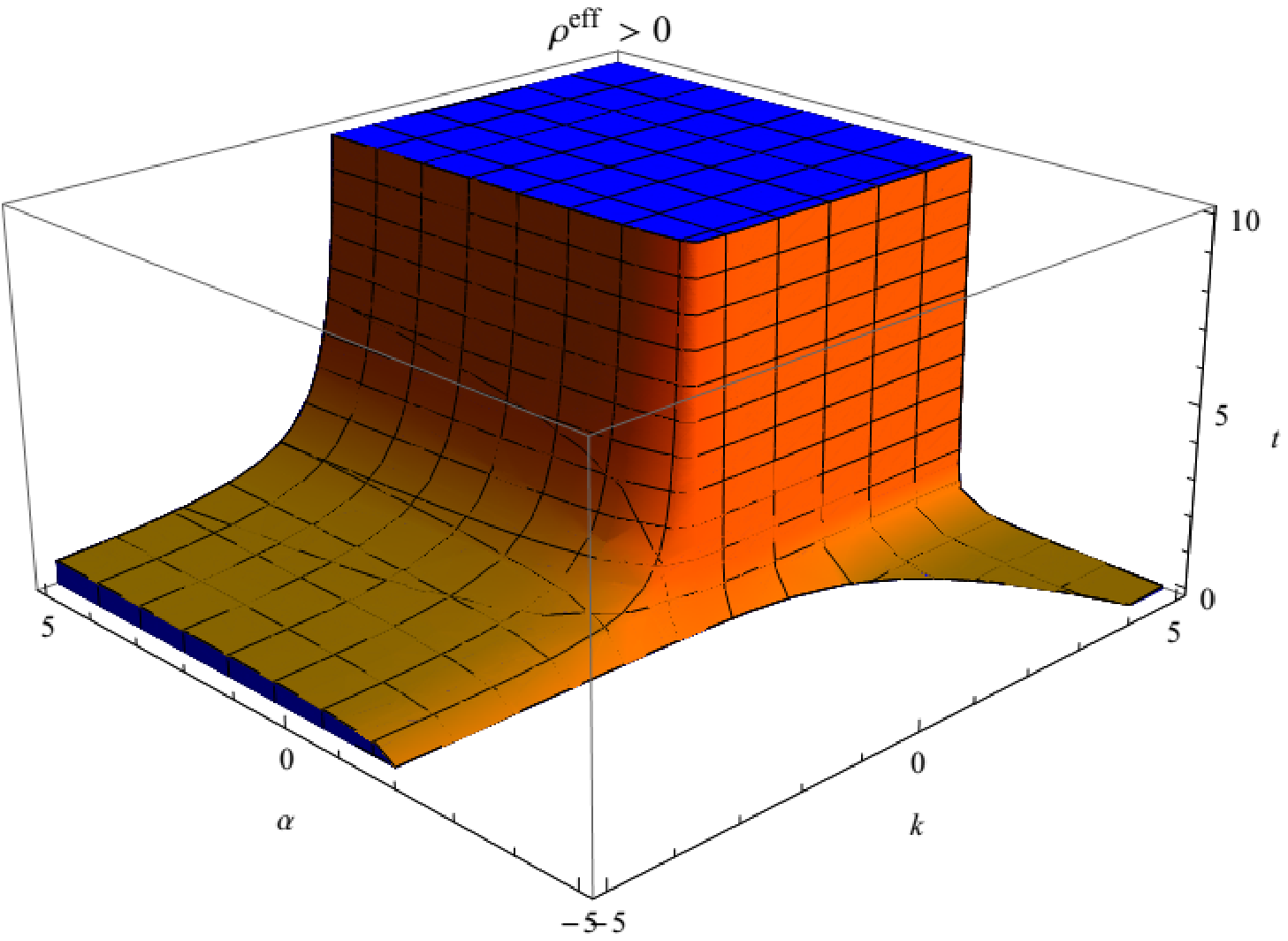,width=.48\linewidth}
\epsfig{file=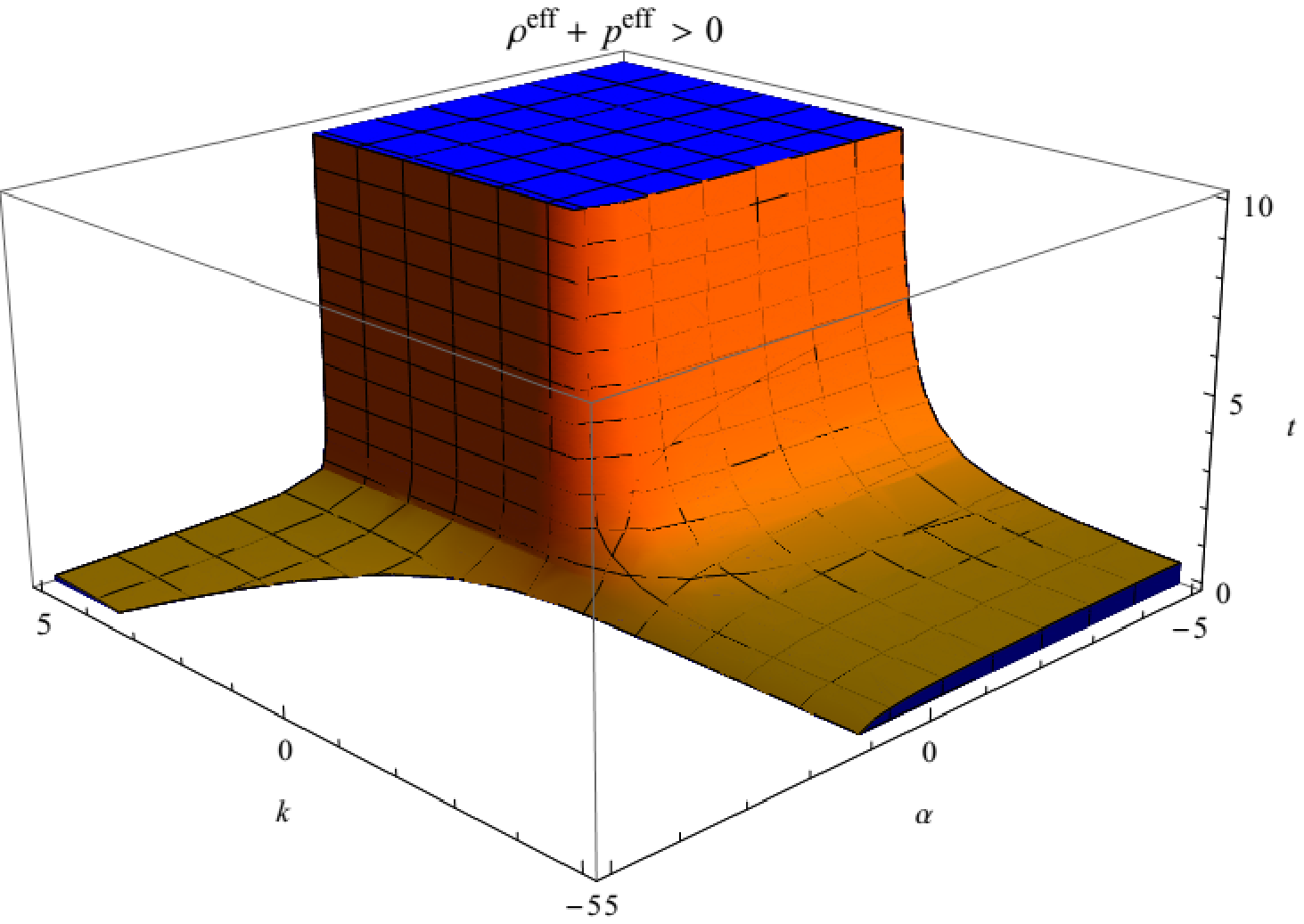,width=.48\linewidth} \caption{WEC plots for model 1 mentioned in Eq.(\ref{model1}) in extended $f(R,G,T)$ gravity. The left plot representing the regions where $\rho_{eff}>0$ while the right plot represents the regions where $\rho_{eff}+ P_{eff}>0$ with respect to $\alpha$, $k$ and $t$, respectively. We see that the WECs is satisfied for the considered range of parameters.} \label{1f2}
\end{figure}

In this context, the plotted region is illustrated in the right diagram of Fig.\ref{1f2}. We concluded that the violation of WEC may be avoided by making all of the parameters ($t,~\alpha,~k$) to be positive in $f(R,G,T) =R+R^2 +\alpha {G^n} + \beta G\log[G]+\gamma T^k$ model.

\begin{itemize}
  \item \textbf{Validity of $\rho^{eff}>0$.}\\

We found that when $t\rightarrow 10$, the desire values of other parameter will get the values of $\alpha\rightarrow  9.92$ and $k\rightarrow -0.179$, while for $t\rightarrow 0.1$ we obtained $\alpha\rightarrow -1.399$ and $k\rightarrow 9.96$.\\

For $\alpha\rightarrow 10$, the other parameter should be $t\rightarrow 0.227$ and $k\rightarrow 9.976$, while for $\alpha\rightarrow -10$ then $t\rightarrow 0.1$ and $k\rightarrow 7.078$.\\

Similarly when $k\rightarrow 10$, the desire values of other parameter will get the values of $\alpha\rightarrow  3.88$ and $t\rightarrow 0.208$ while for $k\rightarrow -10$ we obtained $\alpha\rightarrow -1.673$ and $t\rightarrow 0.126$.\\

\item \textbf{Validity of $\rho^{eff}+P^{eff}>0$.}\\

We found that when $t\rightarrow 10$, the desire values of other parameter will get the values of $\alpha\rightarrow  -1.39$ and $k\rightarrow 9.960$ while for $t\rightarrow 0.1$, we obtained $\alpha\rightarrow -0.896$ and $k\rightarrow 9.354$.\\

For $\alpha\rightarrow 10$, the other parameter should be $t\rightarrow 0.170$ and $k\rightarrow 9.950$, while for $\alpha\rightarrow -10$ then $t\rightarrow 9.47$ and $k\rightarrow -0.05$.\\

Similarly when $k\rightarrow 10$, the desire values of other parameter will get the values of $\alpha\rightarrow  -4.99$ and $t\rightarrow 3.94$, while for $k\rightarrow -10$ we obtained $\alpha\rightarrow -1.673$ and $t\rightarrow 0.126$.\\

\end{itemize}

\subsection{Model 2}

\begin{figure} \centering
\epsfig{file=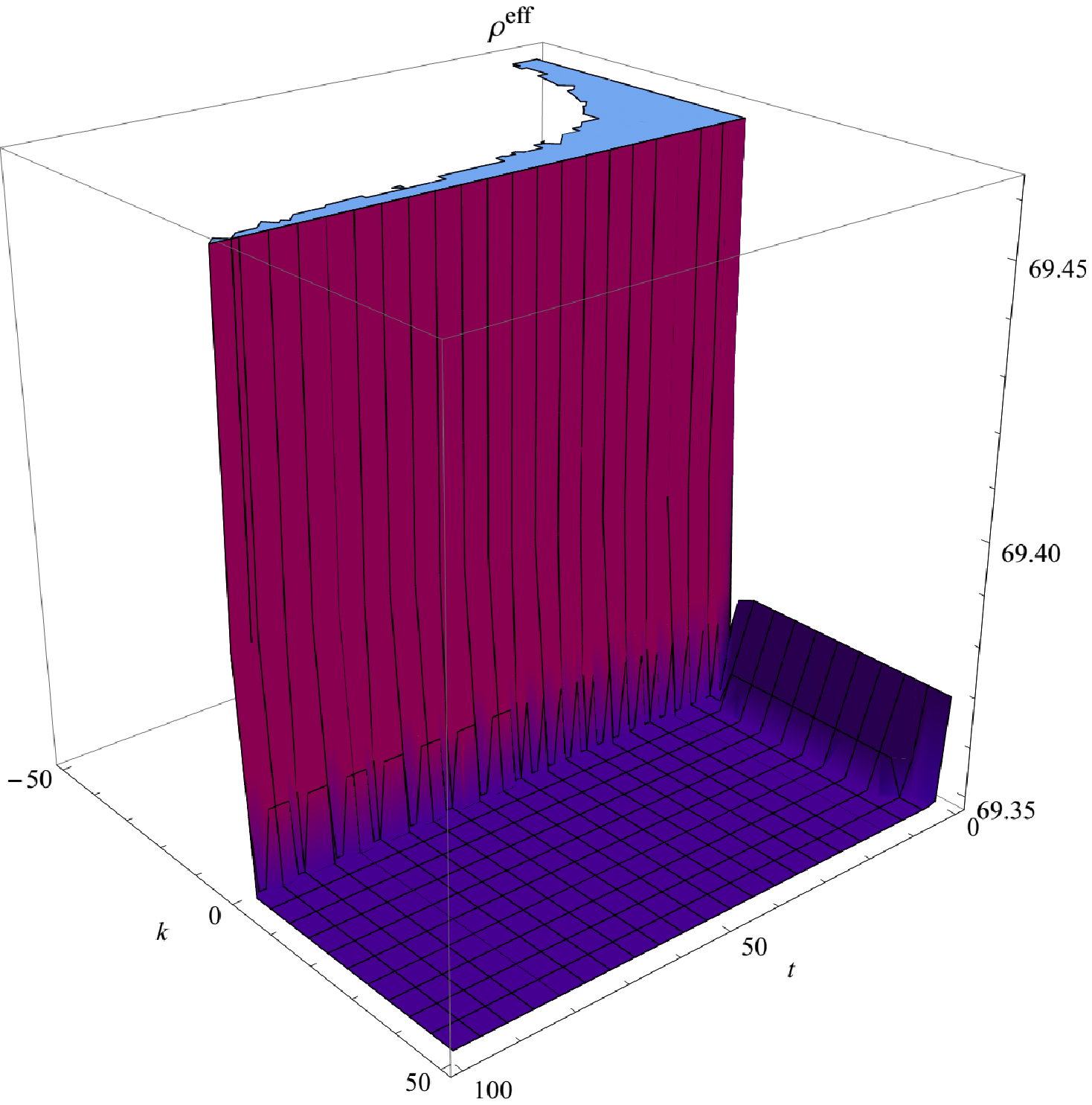,width=.48\linewidth}
\epsfig{file=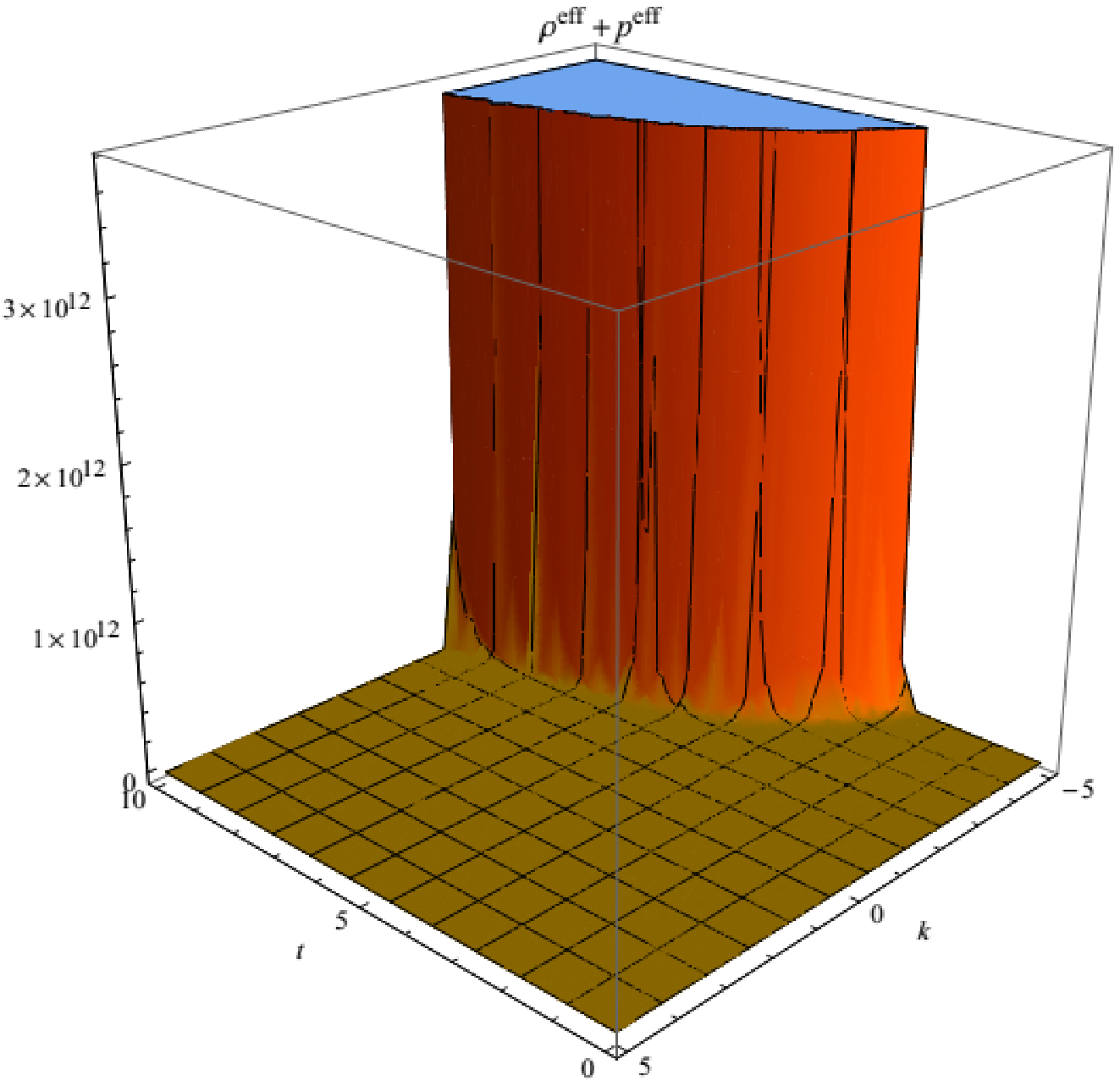,width=.48\linewidth} \caption{WEC plots for model 2 mentioned in Eq.(\ref{model2}) in extended $f(R,G,T)$ gravity. Left plot show the behavior of $\rho^{eff}$ and right plot show the behavior of $\rho^{eff}+ P^{eff}$ with respect to $t$ and $k$ for $\alpha=\beta=1$, respectively.} \label{2f1}
\end{figure}

Next, we look at a more realistic version of the $f(G)$ model \cite{16}, inspire from this, we suggest the following model as
\begin{equation}\label{model2}
f(R,G,T)= R+ R^2+\alpha  G^n \left(\beta  G^m+1\right)+\gamma T^k,
\end{equation}
where $n$ is a positive constant and $\alpha,~\beta$, $k$, $m$ and $\gamma$ are the model parameters. This version might be beneficial for
understanding future singularities in finite time \cite{no08}. Both the local assessment and the galactic limits agree with the conclusions of this hypothesis \cite{no07}.\\
The effective energy density was calculated using Eq.(\ref{model2}) and found to be
\begin{align}\label{rom2}\nonumber
{\rho ^{eff}} &= \frac{1}{{\kappa^2\left( {2R + 1} \right)}}[12(m\alpha \beta {G^{ - 1 + m + n}} + n\alpha {G^{ - 1 + n}}(1 + \beta{G^m}))\\\nonumber
&{H^2}({H^2} - {H^2}(1 + q)) + \frac{1}{2}( - \alpha {G^n}\left( {1 + \beta {G^m}} \right) - R - {R^2} + R\\\nonumber
&\left( {2R + 1} \right) - \gamma {T^k}) + \kappa^2\rho  + k\gamma {T^{ - 1 + k}}\left( {p + \rho } \right)- 12(mn\alpha \beta {G^{ - 2 + m + n}}+m\\
&(m + n - 1)\alpha \beta {G^{ - 2 + m + n}} + \left( {n - 1} \right)n\alpha {G^{ - 2 + n}}\left( {1 + \beta {G^m}} \right)){H^3}G' - 6HR']\geq 0.
\end{align}
While the product of effective pressure and energy density is
\begin{align}\label{roppm2}\nonumber
{\rho ^{eff}} + {P^{eff}} &= \frac{1}{{\left( {{\kappa ^2}{G^3}\left( {1 + 2R} \right)T} \right)}}[4n\left( {2 - 3n + {n^2}} \right)\alpha {G^n}{H^2}T{{G'}^2}\\\nonumber
& + 4({m^3} + 3{m^2}\left( {n - 1} \right) + n(2 - 3n + {n^2}) + m\left( {2 - 6n + 3{n^2}} \right))\alpha \beta {G^{m + n}}\\\nonumber
&{H^2}T{{G'}^2} - 4\left( {n - 1} \right)n\alpha {G^{1 + n}}{H^2}T(H\left( {3 + 2q} \right)G' - {G^{\prime \prime }}) - 4({m^2} + \left( {n - 1} \right)n\\\nonumber
& + m\left( {2n - 1} \right))\alpha \beta {G^{1 + m + n}}{H^2}T(H\left( {2q + 3} \right)G' - {G^{\prime \prime }}) + {G^3}(p({\kappa ^2}T + k\gamma {T^k})\\
& + k\gamma {T^k}\rho  + T({\kappa ^2}\rho  - 2HR' + 2{R^{\prime \prime }}))]\geq 0.
\end{align}
Finding the exact analytical formula for the parameters for these two inequalities  (\ref{rom2}) and (\ref{roppm2}), {are} significantly more difficult.
As a result, we'll set some parameters identical to the specific values. For simplicity, we let $\alpha=1,\beta=1$ and plot $\rho^{\textrm{eff}}$ and $\rho^{\textrm{eff}}+ P^{\textrm{eff}}$, as demonstrated in  Fig.\ref{2f1}. Here we consider $0<t<10$, $-5<k<5$ and $-5<\alpha<5$. WECs is also valid for the model (\ref{model2}) as can be seen in this diagram.
\\

\begin{figure} \centering
\epsfig{file=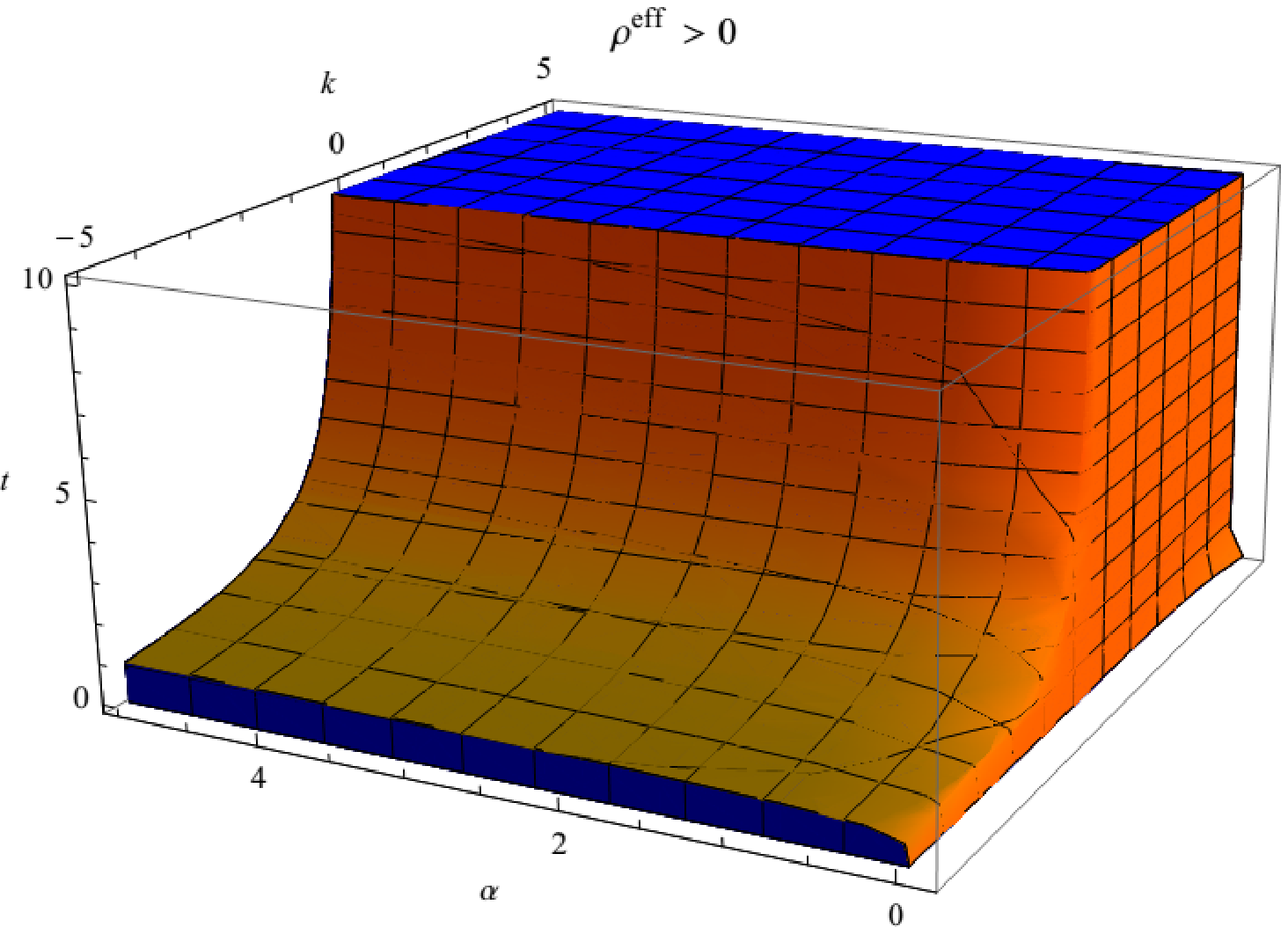,width=.48\linewidth}
\epsfig{file=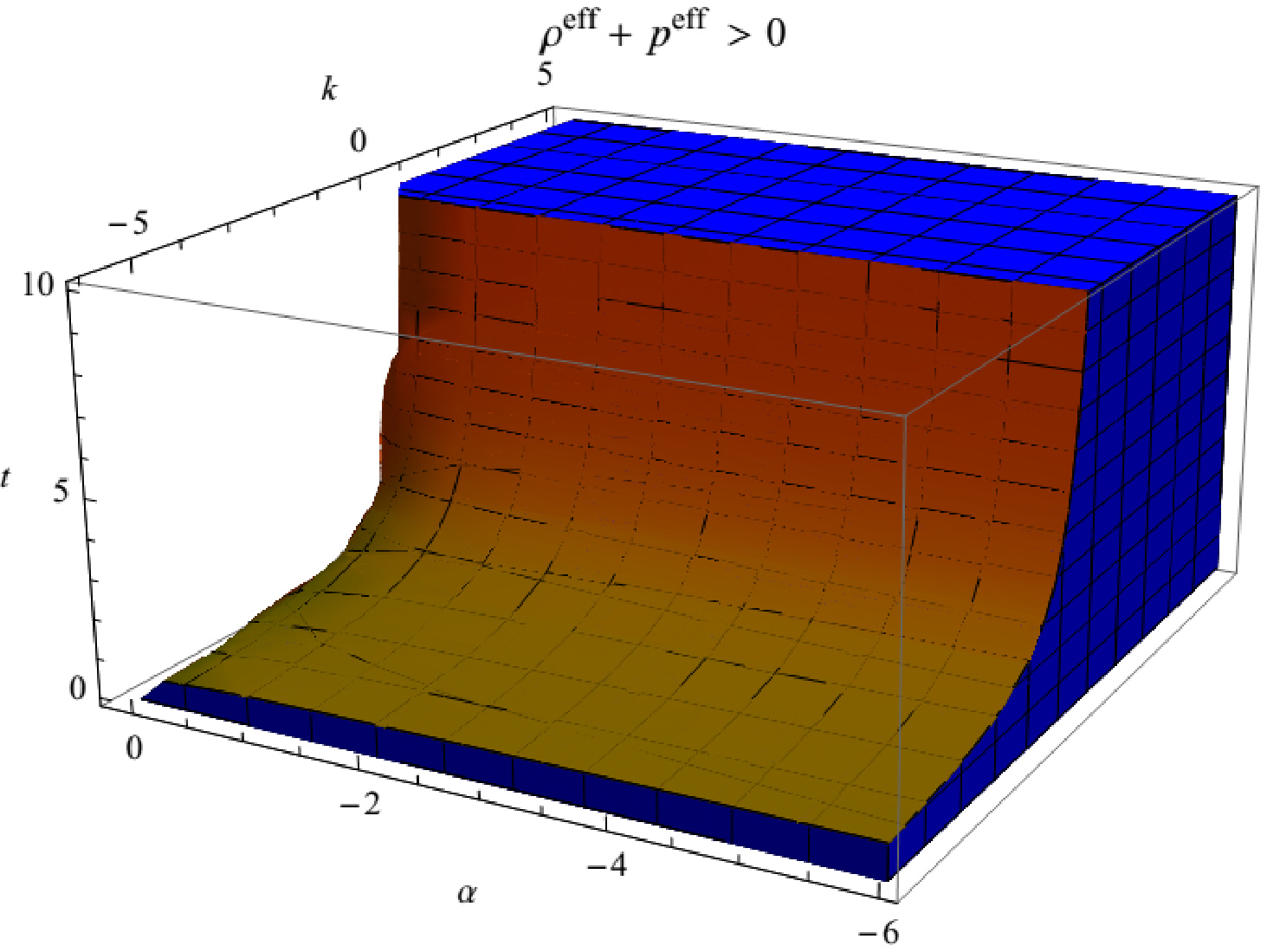,width=.48\linewidth} \caption{WECs plots for model 2 mentioned in Eq.(\ref{model2}) in extended $f(R,G,T)$ gravity. The left plot representing the regions where $\rho_{eff}>0$, while the right plot represents the regions where $\rho_{eff}+ P_{eff}>0$ with respect to $\alpha$, $k$ and $t$, respectively. We see that the WECs is satisfied for the considered range of parameters.}\label{2f2}
\end{figure}

\begin{itemize}
  \item \textbf{Validity of $\rho^{eff}>0$.}\\

We found that when $t\rightarrow 10$, the desire values of other parameter will get the values of $\alpha\rightarrow  9.936$ and $k\rightarrow -0.379$, while for $t\rightarrow 0.1$ we obtained $\alpha\rightarrow 1.080$ and $k\rightarrow 9.957$.\\

For $\alpha\rightarrow 10$, the other parameters should be $t\rightarrow 0.227$ and $k\rightarrow 9.976$, while for $\alpha\rightarrow 0.769$ then $t\rightarrow 0.1$ and $k\rightarrow 10$.\\

Similarly, when $k\rightarrow 10$, the desire values of other parameter will get the values of $\alpha\rightarrow  0.0388$ and $t\rightarrow 0.230$ while for $k\rightarrow -10$, we obtained $\alpha\rightarrow 10$ and $t\rightarrow 0.1$.\\

\textbf{Validity of $\rho^{eff}+P^{eff}>0$.}\\

We found that when $t\rightarrow 10$, the desire values of other parameter will get the values of $\alpha\rightarrow  -9.94$ and $k\rightarrow 5.15$ while for $t\rightarrow 0.1$ we obtained $\alpha\rightarrow -4.75$ and $k\rightarrow -9.78$.\\

For $\alpha\rightarrow 1.17$, the other parameters should be $t\rightarrow 1.17$ and $k\rightarrow 0.1$ while for $\alpha\rightarrow -10$ then $t\rightarrow 7.19$ and $k\rightarrow 9.34$.\\

Similarly, when $k\rightarrow 10$, the desire values of other parameter will get the values of $\alpha\rightarrow  -9.58$ and $t\rightarrow 9.99$ while for $k\rightarrow -10$, we obtained $\alpha\rightarrow 10$ and $t\rightarrow 0.1$.\\

\end{itemize}

\subsection{Model 3}
Some other plausible gravity $f(G)$ model would be fascinating to investigate~\cite{15}, {in this regard}, we suggest the following model as
\begin{equation}\label{model3}
f(R,G,T) = R+ R^2+\frac{{{a_1} {G^n} + {b_1}}}{{{a_2} {G^n} + {b_2}}}+\gamma T^k,
\end{equation}
Here, $a_1,~b_1,~a_2,~b_2$, $k$, $\gamma$ and $n$ are random constants with $n>0$. This model could be beneficial for studying future singularities in early time and late time cosmic acceleration. The effective energy density in this regard
\begin{align}\nonumber
{\rho ^{eff}} & = \frac{1}{{{\kappa ^2}\left( {2R + 1} \right)}}[12( - {a_2}n{G^{ - 1 + n}}\left( {b_1 + a_1{G^n}} \right)\\\nonumber
&{\left( {b_2 + {a_2}{G^n}} \right)^{ - 2}} + a_1 n{\left( {b_2 + a_2{G^n}} \right)^{ - 1}}{G^{ - 1 + n}}){H^2}\left( {{H^2} - {H^2}\left( {1 + q} \right)} \right)\\\nonumber
& + \frac{1}{2}( - (b_1 + a_1{G^n}){\left( {b_2 + a_2{G^n}} \right)^{ - 1}} - R - {R^2} + R\left( {2R + 1} \right) - \gamma {T^k})\\\nonumber
& + {\kappa ^2}\rho  + k\gamma {T^{ - 1 + k}}\left( {p + \rho } \right) - 12(2a_2^2{n^2}{G^{ - 2 + 2n}}\left( {b_1 + a_1{G^n}} \right){\left( {b_2 + a_2{G^n}} \right)^{ - 3}}\\\nonumber
& - 2a_1 a_2{n^2}{\left( {b_2 + a_2{G^n}} \right)^{ - 2}}{G^{ - 2 + 2n}} - a_2\left( {n - 1} \right)n{G^{ - 2 + n}}{\left( {b_2 + a_2{G^n}} \right)^{ - 2}}\\
&\left( {b_1 + a_1{G^n}} \right) + a_1\left( {n - 1} \right){\left( {b_2 + a_2{G^n}} \right)^{ - 1}}n{G^{ - 2 + n}}){H^3}G' - 6HR']\geq 0.
\end{align}
Where the effective energy density and pressure are combined to form
\begin{align}\nonumber
{\rho ^{eff}} + {P^{eff}} &= \frac{1}{{\left\{ {{\kappa ^2}{G^3}{{\left( {b_2 + a_2{G^n}} \right)}^4}\left( {1 + 2R} \right)T} \right\}}}[4b_2^2\left( { - a_2 b_1 + a_1 b_2} \right)\\\nonumber
&n\left( {{n^2} - 3n + 2} \right){G^n}{H^2}T{{G'}^2} + 16a_2 b_2\left( {a_2 b_1 - a_1 b_2} \right)n\left( {{n^2} - 1} \right){G^{2n}}{H^2}T{{G'}^2}\\\nonumber
& - 4a_2^2\left( {a_2 b_1 - a_1 b_2} \right)n\left( {{n^2} + 3n + 2} \right){G^{3n}}{H^2}T{{G'}^2} + 4b_2^2\left( {a_2 b_1 - a_1 b_2} \right)\\\nonumber
&\left( {n - 1} \right)n{G^{1 + n}}{H^2}T(H\left( {3 + 2q} \right)G' - {G^{\prime \prime }}) - 8a_2 b_2\left( {a_2 b_1 - a_1 b_2} \right)n{G^{1 + 2n}}{H^2}T\\\nonumber
&\left( {H\left( {3 + 2q} \right)G' - {G^{\prime \prime }}} \right) - 4a_2^2\left( {a_2 b_1 - a_1 b_2} \right)n\left( {1 + n} \right){G^{1 + 3n}}{H^2}T(H\left( {3 + 2q} \right)G'\\\nonumber
& - {G^{\prime \prime }}) + b_2^4{G^3}\left( {p\left( {{\kappa ^2}T + k\gamma {T^k}} \right) + k\gamma {T^k}\rho  + T\left( {{\kappa ^2}\rho  - 2H R' + 2{R^{\prime \prime }}} \right)} \right) + 4a_2 b_2^3\\\nonumber
&{G^{3 + n}}(p\left( {{\kappa ^2}T + k\gamma {T^k}} \right) + k\gamma {T^k}\rho  + T\left( {{\kappa ^2}\rho  - 2H R' + 2{R^{\prime \prime }}} \right)) + 6a_2^2b_2^2{G^{3 + 2n}}\\\nonumber
&\left( {p\left( {{\kappa ^2}T + k\gamma {T^k}} \right) + k\gamma {T^k}\rho  + T\left( {{\kappa ^2}\rho  - 2H R' + 2{R^{\prime \prime }}} \right)} \right) + 4a_2^3b_2{G^{3 + 3n}}\\\nonumber
&(p\left( {{\kappa ^2}T + k\gamma {T^k}} \right) + k\gamma {T^k}\rho  + T\left( {{\kappa ^2}\rho  - 2H R' + 2{R^{\prime \prime }}} \right)) + a_2^4{G^{3 + 4n}}(p\\
&\left( {{\kappa ^2}T + k\gamma {T^k}} \right) + k\gamma {T^k}\rho  + T\left( {{\kappa ^2}\rho  - 2H R' + 2{R^{\prime \prime }}} \right))]\geq 0.
\end{align}
This model having variables, i.e., $a_1,~a_2,~b_1,~b_2,~m$, so we'll assign few specific values to some of these parameters to constraint them with a specific range. For the sake of simplicity, we'll set $b_1=-1, b_2=1$. Here we consider $0<t<10$, $-5<k<5$ and $-5<\alpha<5$. Other attributes are:

\begin{figure} \centering
\epsfig{file=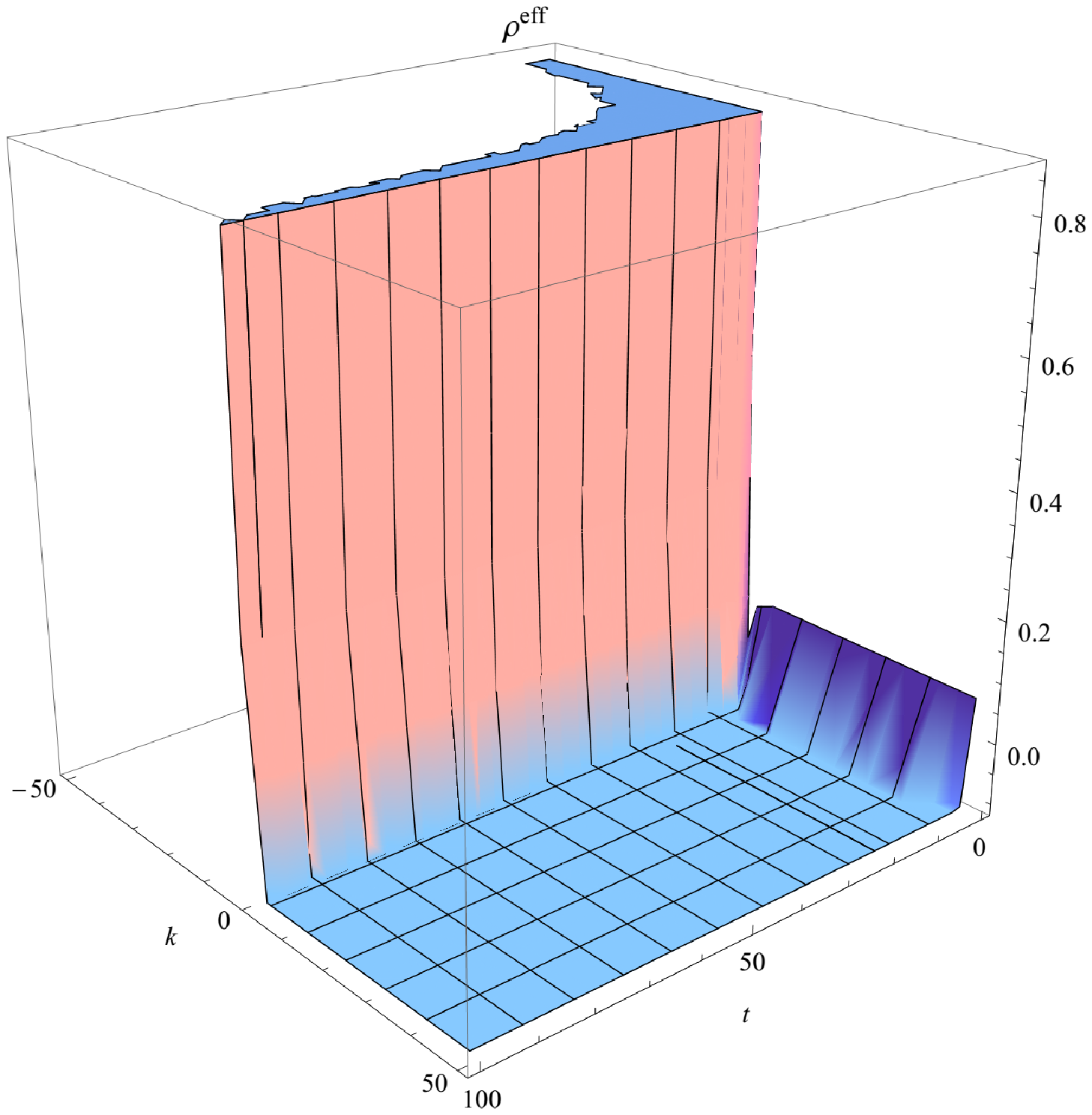,width=.48\linewidth}
\epsfig{file=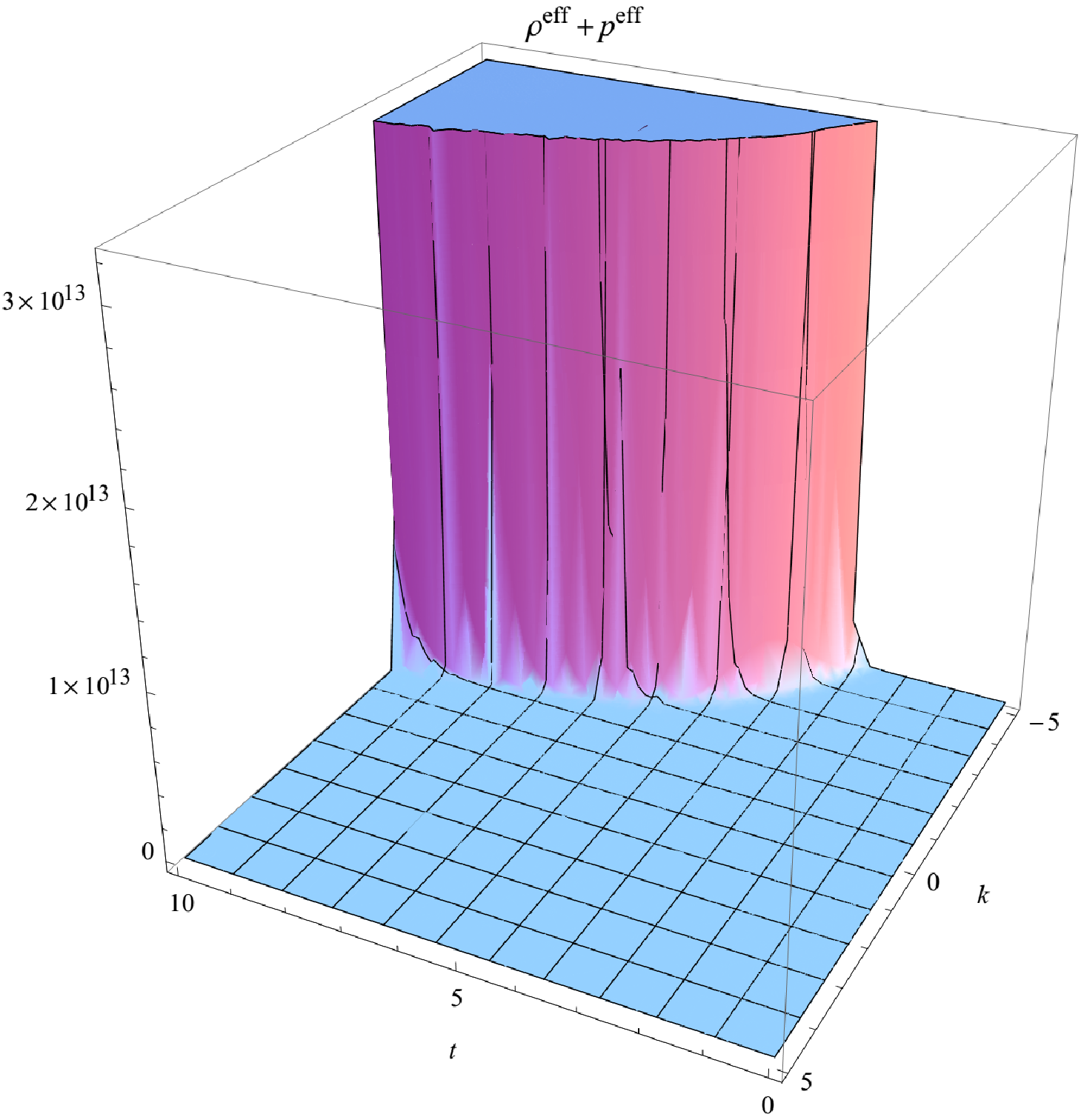,width=.48\linewidth} \caption{WEC plots for model 3 mentioned in Eq.(\ref{model3}) in extended $f(R,G,T)$ gravity. Left plot show the behavior of $\rho^{eff}$ and right plot show the behavior of $\rho^{eff}+ P^{eff}$ with respect to $t$ and $k$ for $\alpha=\beta=1$, respectively.} \label{3f1}
\end{figure}

\begin{figure} \centering
\epsfig{file=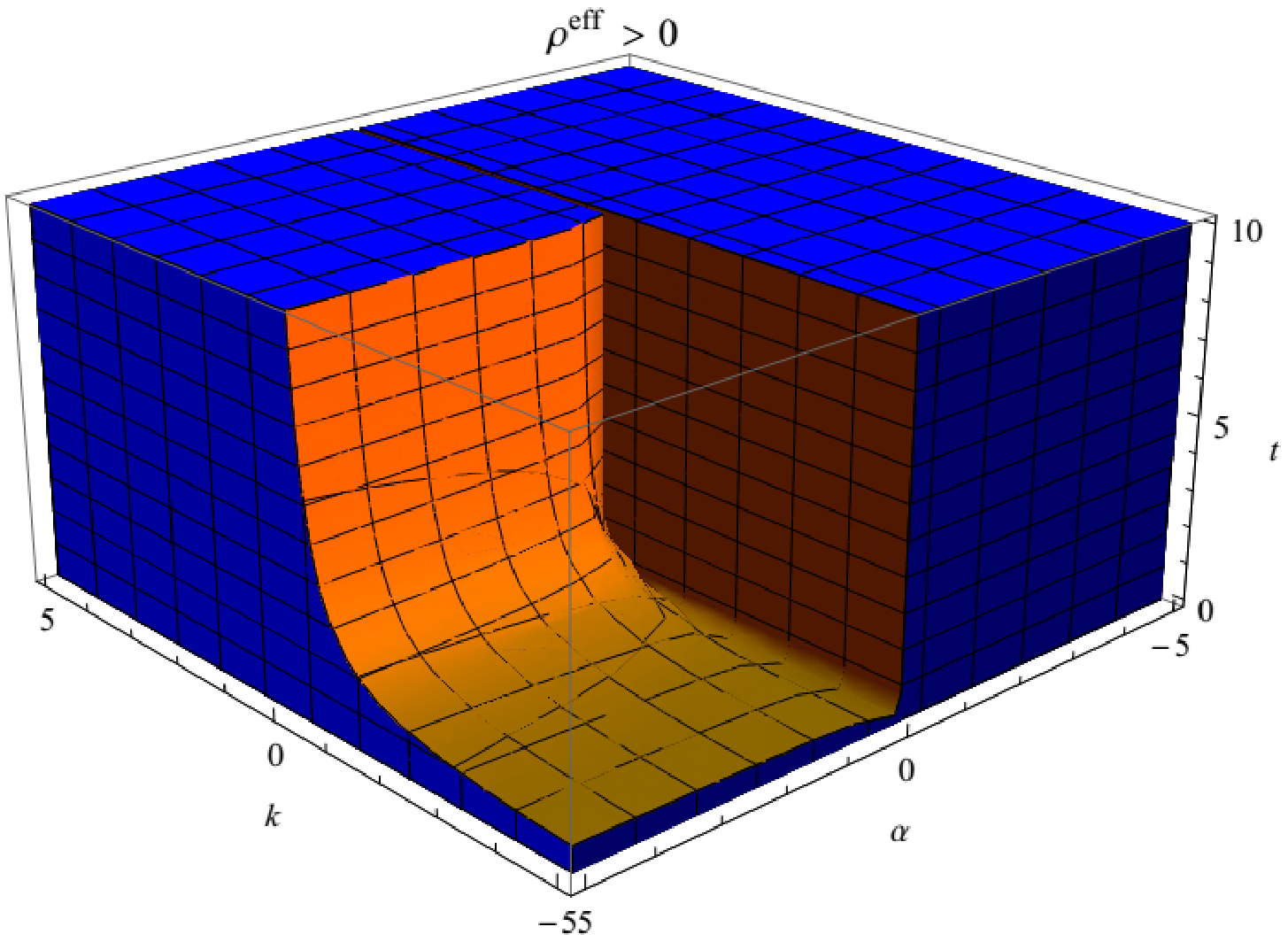,width=.48\linewidth}
\epsfig{file=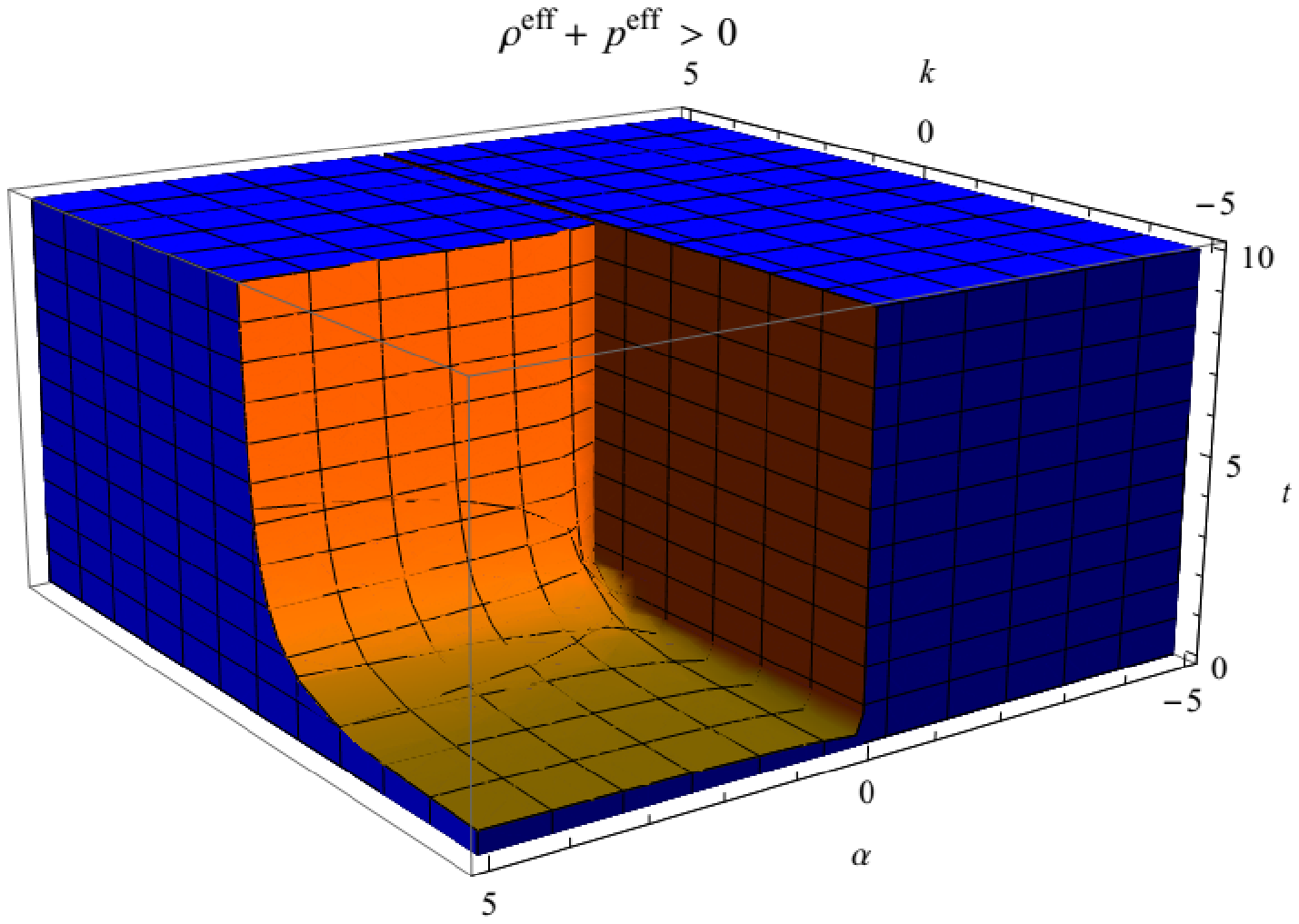,width=.48\linewidth} \caption{WECs plots for model 3 mentioned in Eq.(\ref{model3}) in extended $f(R,G,T)$ gravity. The left plot representing the regions where $\rho_{eff}>0$ while the right plot represents the regions where $\rho_{eff}+ P_{eff}>0$ with respect to $\alpha$, $k$ and $t$, respectively. We see that the WECs is satisfied for the considered range of parameters.} \label{3f2}
\end{figure}

\begin{itemize}
  \item \textbf{Validity of $\rho^{eff}>0$.}\\

We found that when $t\rightarrow 10$, the desire values of other parameter will get the values of $\alpha\rightarrow  3.077$ and $k\rightarrow 9.894$, while for $t\rightarrow 0.1$ we obtained $\alpha\rightarrow -9.093$ and $k\rightarrow -3.349$.\\

For $\alpha\rightarrow 10$, the other parameter should be $t\rightarrow 9.942$ and $k\rightarrow 8.316$ while for $\alpha\rightarrow -10$ then $t\rightarrow 0.137$ and $k\rightarrow -9.99$.\\

Similarly, when $k\rightarrow 10$, the desire values of other parameter will get the values of $\alpha\rightarrow  2.151$ and $t\rightarrow 9.64$ while for $k\rightarrow -10$ we obtained $\alpha\rightarrow -7.29$ and $t\rightarrow 0.987$.\\

  \item \textbf{Validity of $\rho^{eff}+P^{eff}>0$.}\\

We found that when $t\rightarrow 10$, the desire values of other parameter will get the values of $\alpha\rightarrow  -3.077$ and $k\rightarrow 8.316$ while for $t\rightarrow 0.1$ we obtained $\alpha\rightarrow -6.01$ and $k\rightarrow -9.768$.\\

For $\alpha\rightarrow 1.17$, the other parameter should be $t\rightarrow 9.942$ and $k\rightarrow 8.316$ while for $\alpha\rightarrow -10$ then $t\rightarrow 0.137$ and $k\rightarrow -9.99$.\\

Similarly, when $k\rightarrow 10$, the desire values of other parameter will get the values of $\alpha\rightarrow  2.15$ and $t\rightarrow 9.64$ while for $k\rightarrow -10$. {We} obtained $\alpha\rightarrow -7.29$ and $t\rightarrow 0.98$.\\

\end{itemize}
\section{Stability of Models}
It is crucial to make an argument for the stability of the suggested models in the $f(R,G,T)$ gravity theory here. This universe can be thought of as a thermodynamical system given that it is made up entirely of perfect fluid. This goal is accomplished by introducing an arbitrary function (the quantity of sound speed ($c_s^2$)) to describe the system made up of the ideal fluid. This arbitrary function can be expressed in terms of the universe's effective energy density ($\rho_{eff}$) and effective pressure ($P_{eff}$ ) as follows:

$$c_{s}^2= \frac{dP_{eff}}{d\rho_{eff}}.$$

\begin{figure} \centering
\epsfig{file=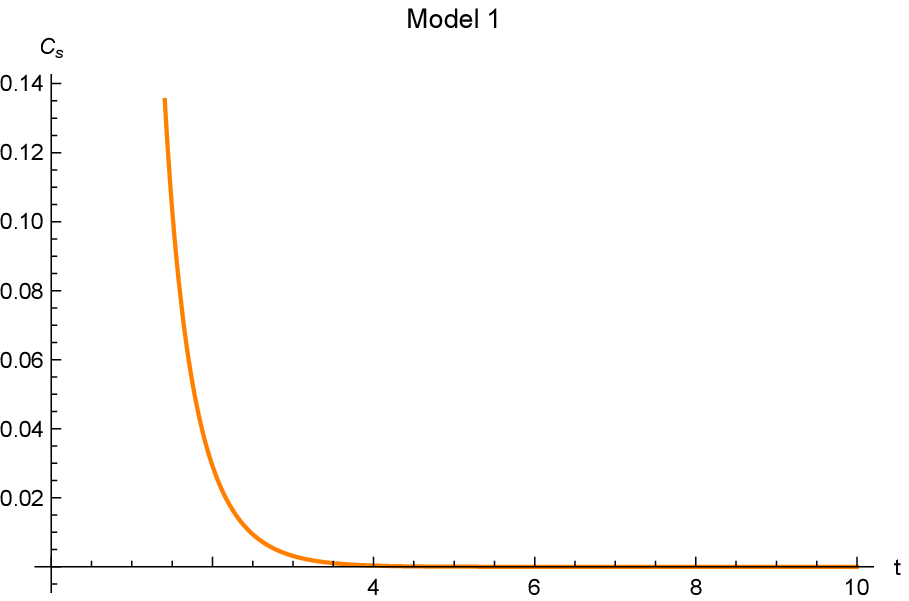,width=.4\linewidth}
\epsfig{file=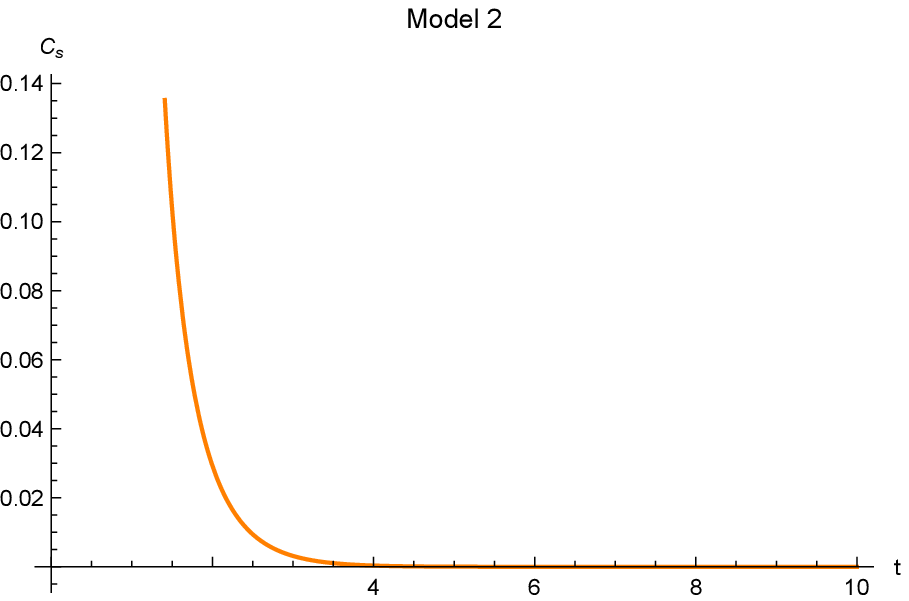,width=.4\linewidth}
\epsfig{file=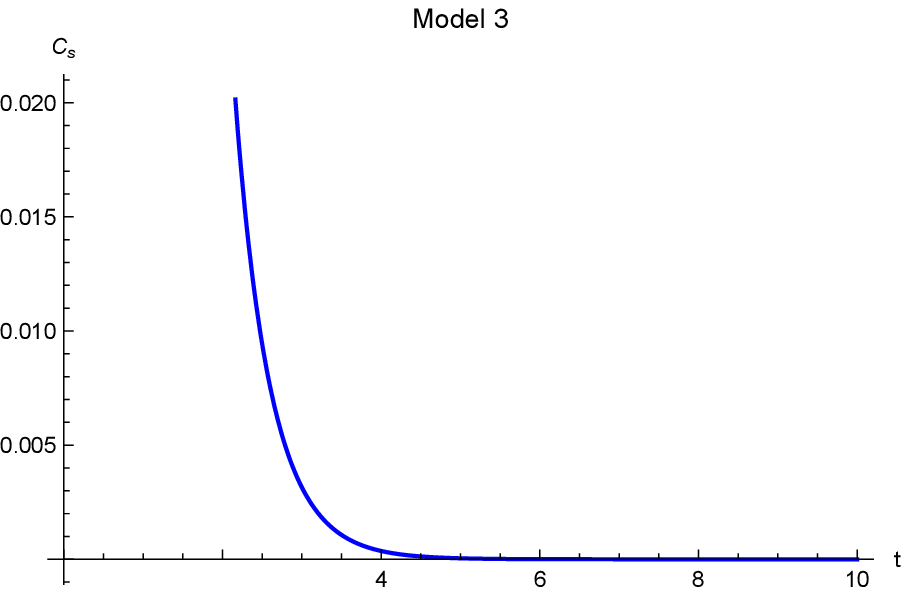,width=.4\linewidth} \caption{Stability of Models} \label{3f3}
\end{figure}

It is known that the value of the aforementioned function is greater than zero for a thermodynamic system. In light of this, stability conditions can be verified when $c_s^2>0$. Using adiabatic and non-adiabatic methods, a thermodynamic system can be explained perturbatively. Effective energy density, effective pressure, and universe entropy are three potential disturbance quantities in this situation. The stability of our models can plotted in Fig. \ref{3f3}.

\section{Summary}

\vspace{0.5cm}

We investigated the impact of enhanced extended theory of gravity models on the occurrence of actual cosmological perfect fluid configurations. The study of ECs are inextricably linked to a pragmatic representation of WH solutions \cite{i1,i2,i3,i4,i5,i6}. Inspection of plausible and quite well models is an enticing goal for avoiding the utilization of unusual matter content at the WH throat. We looked at the behaviour of the FLRW metric when it was full with an perfect fluid. The field equations $f(R,G,T)$ turn out to be extremely nonlinear, and Even without any fundamentally coherent principles, they can't be solved. We calculated the general energy inequalities relationship using $f(R,G,T)$ field equations. Three separate customized extended theory of gravity simulations were examined, i.e., $f(R,G,T) =R+R^2 +\alpha {G^n} + \beta G\log[G]+\gamma T^k,~f(R,G,T)=R+R^2+ \alpha  G^n \left(\beta G^m+1\right)+\gamma T^k$ and $f(R,G,T) =R+R^2+ \frac{{{a_1} {G^n} + {b_1}}}{{{a_2} {G^n} + {b_2}}}+\gamma T^k$. We examined the behaviour of ECs using all of the customized extended models discussed over.
Then, with the several possible $f(R,G,T)$ models, the most likely measured results of the parameters; Hubble, deceleration, jerk, and snap are employed. The areas where NEC and WEC maintain under different $f(R,G,T)$ gravity values were displayed. Here, we consider $0<t<10$, $-5<k<5$ and $-5<\alpha<5$. These following are several of the graphical features that demonstrate some of the results:\\
\textbf{Bounds on Model 1:}
\begin{itemize}
  \item For time, $0<t<1$, $\rho>0$ for all $k$, along with $\alpha>-1.7$.
  \item For parameter $k$, $4<k<5$, $\rho>0$ for all $\alpha$, along with a very small $t$.
  \item Furthermore, $\rho>0$ for $t>0$, along with $k<0$ and $\alpha>-1.7$.
\end{itemize}
\textbf{Bounds on Model 2:}
\begin{itemize}
 \item For small value of (e.g., $0<t<1$), the density, $\rho>0$ for any value of $k$, along with positive value of $\alpha>0$.
\item For $t>1$, the validity of $\rho>0$ required both $\alpha>0$ with $k>0$.
\item For small value of (e.g., $0<t<1$), the density, $\rho+P>0$ for any value of $k$, along with positive value of $\alpha<0$.
\item For $t>1$, the validity of $\rho+P>0$ required both $\alpha<0$ with $k>0$.
\end{itemize}
\textbf{Bounds on Model 3:}
\begin{itemize}
\item  For $0<t<1$, the validity of WEC required the parameter $k$ and $\alpha$ to be any value in given range.
\item  For $t>1$, WEC bounds the parameters $\alpha<0$, along with any value of $k$.
\item  Similarly, for $t>1$, WEC also restrict the parameters $\alpha>0$, along with any value of $k>0$.
\end{itemize}

The suggested models’ solutions found in this research will be useful for creating a physical and realistic model that accounts for the universe's acceleration. When compared to the discussion of pure gravity, theoretical study may then produce some qualitative outcomes. It will be used somewhere else.\\
It is important to look into the possibility of our models’ solutions for various cosmological phases. Because they exist in a FLRW background that reflects all potential cosmic evolutions, including the dark energy period, matter dominant, or radiation dominant eras, these solutions are especially significant.


\vspace{0.3cm}

\bibliographystyle{unsrt}
\bibliography{mybib}

\end{document}